\documentstyle[11pt]{article}

\expandafter\ifx\csname amssym.def\endcsname\relax \else\endinput\fi
\expandafter\edef\csname amssym.def\endcsname{%
       \catcode`\noexpand\@=\the\catcode`\@\space}
\catcode`\@=11
\def\undefine#1{\let#1\undefined}
\def\newsymbol#1#2#3#4#5{\let\next@\relax
 \ifnum#2=\@ne\let\next@\msafam@\else
 \ifnum#2=\tw@\let\next@\msbfam@\fi\fi
 \mathchardef#1="#3\next@#4#5}
\def\mathhexbox@#1#2#3{\relax
 \ifmmode\mathpalette{}{\m@th\mathchar"#1#2#3}%
 \else\leavevmode\hbox{$\m@th\mathchar"#1#2#3$}\fi}
\def\hexnumber@#1{\ifcase#1 0\or 1\or 2\or 3\or 4\or 5\or 6\or 7\or 8\or
 9\or A\or B\or C\or D\or E\or F\fi}

\font\tenmsa=msam10
\font\sevenmsa=msam7
\font\fivemsa=msam5
\newfam\msafam
\textfont\msafam=\tenmsa
\scriptfont\msafam=\sevenmsa
\scriptscriptfont\msafam=\fivemsa
\edef\msafam@{\hexnumber@\msafam}
\mathchardef\dabar@"0\msafam@39
\def\dashrightarrow{\mathrel{\dabar@\dabar@\mathchar"0\msafam@4B}}
\def\dashleftarrow{\mathrel{\mathchar"0\msafam@4C\dabar@\dabar@}}

\def\ulcorner{\delimiter"4\msafam@70\msafam@70 }
\def\urcorner{\delimiter"5\msafam@71\msafam@71 }
\def\llcorner{\delimiter"4\msafam@78\msafam@78 }
\def\lrcorner{\delimiter"5\msafam@79\msafam@79 }
\def\yen{{\mathhexbox@\msafam@55}}
\def\checkmark{{\mathhexbox@\msafam@58}}
\def\circledR{{\mathhexbox@\msafam@72}}
\def\maltese{{\mathhexbox@\msafam@7A}}
\def\circledS{{\mathhexbox@\msafam@73}}

\font\tenmsb=msbm10
\font\sevenmsb=msbm7
\font\fivemsb=msbm5
\newfam\msbfam
\textfont\msbfam=\tenmsb
\scriptfont\msbfam=\sevenmsb
\scriptscriptfont\msbfam=\fivemsb
\edef\msbfam@{\hexnumber@\msbfam}
\def\Bbb#1{{\fam\msbfam\relax#1}}
\def\widehat#1{\setbox\z@\hbox{$\m@th#1$}%
 \ifdim\wd\z@>\tw@ em\mathaccent"0\msbfam@5B{#1}%
 \else\mathaccent"0362{#1}\fi}
\def\widetilde#1{\setbox\z@\hbox{$\m@th#1$}%
 \ifdim\wd\z@>\tw@ em\mathaccent"0\msbfam@5D{#1}%
 \else\mathaccent"0365{#1}\fi}
\font\teneufm=eufm10
\font\seveneufm=eufm7
\font\fiveeufm=eufm5
\newfam\eufmfam
\textfont\eufmfam=\teneufm
\scriptfont\eufmfam=\seveneufm
\scriptscriptfont\eufmfam=\fiveeufm

\csname amssym.def\endcsname

\makeatletter
\def\section{\@startsection {section}{1}{\z@}{-3.5ex plus -1ex minus 
 -.2ex}{2.3ex plus .2ex}{\large\sc}}
\def\subsection{\@startsection{subsection}{2}{\z@}{-3.25ex plus -1ex minus 
 -.2ex}{1.5ex plus .2ex}{\normalsize\sc}}
\makeatother

\makeatletter
\@addtoreset{equation}{section}                        

\makeatother

\newcommand{\nc}{\newcommand}
\newcommand{\rnc}{\renewcommand}

\nc{\be}{\begin{equation}}
\nc{\ee}{\end{equation}}
\nc{\bea}{\begin{eqnarray}}
\nc{\eea}{\end{eqnarray}}

\nc{\trac}[2]{{\textstyle\frac{#1}{#2}}}

\nc{\ex}[1]{\mbox{e}^{\,\textstyle#1}}

\nc{\CC}{\Bbb{C}}
\nc{\HH}{\Bbb{H}}
\nc{\PP}{\Bbb{P}}
\nc{\RR}{\Bbb{R}}
\nc{\ZZ}{\Bbb{Z}}
\nc{\II}{\Bbb{I}}
\nc{\EE}{\Bbb{E}}
\nc{\SS}{\Bbb{S}}

\rnc{\a}{\alpha}
\nc{\al}{\a^{l}}
\rnc{\d}{\delta}
\nc{\ga}{\gamma}
\nc{\la}{\lambda}
\nc{\lal}{\la_{l}}
\nc{\f}{\phi}
\nc{\fb}{\bar{\phi}}
\nc{\p}{\psi}

\nc{\e}{\eta}
\nc{\eb}{\bar{\eta}}
\rnc{\c}{\chi}
\nc{\eps}{\epsilon}
\rnc{\t}{\theta}
\nc{\tb}{\bar{\theta}}
\nc{\om}{\omega}

\rnc{\P}{\Psi}
\nc{\pl}{\P_{L}}
\nc{\pdr}{\P^{\dag}_{R}}
\nc{\G}{\Gamma}

\nc{\sig}{\sigma}
\nc{\sk}{\sigma_{k}}
\nc{\sa}{\sigma_{a}}
\nc{\Bb}{\bar{B}}

\nc{\symx}{\circledS}

\nc{\Q}{\bar{Q}}
\nc{\C}{{\cal A}/{\cal G}}                
\nc{\A}[1]{{\cal A}^{#1}/{\cal G}^{#1}}  
\nc{\RC}{{\cal R}_{\C}}                 
\nc{\RM}{{\cal R}_{\M}}                
\nc{\RX}{{\cal R}_{X}}
\nc{\RY}{{\cal R}_{Y}}

\nc{\ad}{\mathop{\mbox{ad}}\nolimits}
\nc{\tr}{\mathop{\mbox{tr}}\nolimits}
\nc{\Tr}{\mathop{\mbox{Tr}}\nolimits}
\nc{\Det}{\mathop{\mbox{Det}}\nolimits}
\rnc{\det}{\mathop{\mbox{det}}\nolimits}
\nc{\rk}{\mathop{\mbox{rk}}\nolimits}
\nc{\diag}{\mbox{diag}}
\nc{\ra}{\rightarrow}
\nc{\Ra}{\Rightarrow}
\nc{\LRa}{\Leftrightarrow}
\nc{\lra}{\leftrightarrow}
\nc{\ot}{\otimes}
\rnc{\ss}{\subset}
\nc{\nul}{\noindent\underline}
\nc{\non}{\nonumber\\}
\rnc{\S}{\Sigma}
\nc{\tp}{2\pi i}
\nc{\del}{\partial}
\nc{\dbar}{\bar{\del}}
\nc{\dx}{\dot{x}}
\nc{\zb}{\bar{z}}

\rnc{\lg}{\log g^{2}}
\nc{\lv}{\log V_{s}}
\nc{\vs}{V_{s}}
\rnc{\ln}{\log \N}
\nc{\ls}{\ell_{s}}
\nc{\N}{{\cal N}}
\nc{\M}{{\cal M}}
\nc{\F}{{\cal F}}
\nc{\E}{{\cal E}}
\rnc{\P}{{\cal P}}
\nc{\I}{{\cal I}}
\nc{\IIt}{$\widetilde{\mbox{II}}$}
\nc{\gst}{\widetilde{g_{s}}}
\nc{\gsh}{\widehat{g_{s}}}
\nc{\lsh}{\widehat{\ls}}
\nc{\rllh}{\widehat{R_{11}}}
\nc{\lph}{\widehat{\ell_{P}}}
\nc{\mnd}{M_{Nd}}

\nc{\mat}[4]{\left(\begin{array}{cc}#1&#2\\#3&#4\end{array}\right)}
 
\nc{\r}[1]{\mathbf{#1}}
\nc{\rb}[1]{\overline{\mathbf{#1}}}

\nc{\gi}{\gamma_{i}}
\nc{\gj}{\gamma_{j}}

\nc{\subs}[1]{{\vspace*{0.5cm}}%
{\noindent\underline{\small\sc #1}}{\addcontentsline{toc}{subsubsection}{#1}}%
{\vspace*{0.3cm}}}

\nc{\chap}[1]{{\clearpage}%
\begin{center}%
{\noindent\underline{\large\sc #1}}{\addcontentsline{toc}{section}{#1}}%
\end{center}%
{\vspace*{0.3cm}}}

\newcommand{\ba}{\begin{eqnarray}}
\newcommand{\ea}{\end{eqnarray}}
\newcommand{\rarw}{\rightarrow}
\newcommand{\lrarw}{\leftrightarrow}

\begin{document}
\begin{titlepage}

\begin{flushright}
IC/97/198\\hep-th/9712047\\
\end{flushright}
\vspace*{0.5in}
\begin{center}

{\LARGE{\sc Aspects of U-Duality in Matrix Theory}}\\
\vskip .3in
{\large\sc Matthias Blau} and 
{\sc Martin O'Loughlin}\footnote{e-mail: mblau,mjol@ictp.trieste.it}\\
\vspace{.2in}
{\it ICTP\\ Strada Costiera 11, 34014 Trieste\\ Italy}
\end{center}

\begin{abstract}
\noindent We explore various aspects of implementing the full M-theory
U-duality group $E_{d+1}$, and thus Lorentz invariance, in
the finite $N$ matrix theory (DLCQ of M-theory) describing 
toroidal IIA-compactifications on $d$-tori: (1) We generalize
the analysis of Elitzur et al.\ (hep-th/9707217) from $E_{d}$
to $E_{d+1}$ and identify the highest weight states unifying
the momentum and flux $E_{d}$-multiplets into one $E_{d+1}$-orbit.
(2) We identify the new symmetries, in particular the Weyl group
symmetry associated to the $(d+1)$'th node of the $E_{d+1}$
Dynkin diagram, with Nahm-duality-like symmetries (N-duality)
exchanging the rank $N$ of the matrix theory gauge group with
other (electric, magnetic, \ldots) quantum numbers. (3) We
describe the action of N-duality on BPS bound states, thus making 
testable predictions for the Lorentz invariance of matrix 
theory. (4) We discuss the problems that arise in the matrix
theory limit for BPS states with no top-dimensional branes, i.e.\
configurations with $N = 0$.
(5) We show that N-duality maps the matrix theory SYM picture 
to the matrix string picture and argue that, for $d$ even, the 
latter should be thought of as an M-theory membrane description
(which appears to be well defined even for $d>5$). 
(6) We find a compact and unified expression for a U-duality
invariant of $E_{d+1}$ for all $d$ and show that in $d=5,6$
it reduces to the black hole entropy 
cubic $E_{6}$- and quartic $E_{7}$-invariants
respectively.
(7) Finally, we describe some of the solitonic states in $d=6,7$
and give an example (a `rolled-up' Taub-NUT 6-brane) of a
configuration exhibiting the unusual $1/g_{s}^{3}$-behaviour.
\end{abstract}
\end{titlepage}
\setcounter{footnote}{0}
\begin{small}
\tableofcontents
\end{small}

\section{Introduction}

It is well known that the equations of motion of
eleven-dimensional supergravity compactified on a 
$(d+1)$-torus exhibit a hidden non-compact
global symmetry group $E_{d+1(d+1)}$.\footnote{$E_{3(3)} = 
SL(3) \times SL(2),\, E_{4(4)} = SL(5),\, 
E_{5(5)} = SO(5,5)$} Once the massive
modes of string theory are included, this
full structure does not survive. However, in \cite{hulltown} it 
was conjectured that a discrete subgroup $E_{d+1(d+1)}(\ZZ)$ 
survives in the full string theory, in particular via its action on the
BPS spectrum and as a discrete set of identifications on the supergravity 
moduli space. This group, known as the U-duality group, has played a
central role in the subsequent investigations of string theory dualities.
 
In a recent paper \cite{egkr},
the action of the Weyl group of the corresponding $E_d \subset E_{d+1}$
upon states arising from wrapped M-branes was discussed. These results
taken together with the dictionary from M-theory to the super-Yang-Mills
(SYM) variables
of matrix theory \cite{bfss}
enable one to discuss the action of U-duality in 
the SYM variables, and hence on the BPS spectrum of matrix theory. 
An excellent and authoritative review of matrix theory is now available
\cite{tbreview}, and we refer to it for an extensive list of references
to earlier work on various aspects of matrix theory.

As the eleventh direction of M-theory plays a distinguished role in the
transition from M-theory to its matrix theory description, it at first
does not appear straightforward to extend this to an action of the 
full duality group $E_{d+1}$. Here we will propose such an extension,
based once again on the M-theory dictionary as well as on recent 
observations by Sen and Seiberg \cite{sennew,seibergnew} connecting
the matrix theory limit of M-theory to the discrete light-cone quantization
(DLCQ) of M-theory proposed by Susskind \cite{susskind}. 

A small but crucial difference 
between our approach and that of \cite{egkr} is that we consider the
action of the Weyl group on the momenta of the BPS states, rather 
than on the energies. This turns out to be essential in the SYM picture of 
compactified matrix theory, if one wishes to combine the
momentum and flux multiplets into a single multiplet of
the full U-duality group. 

The issue of finding the matrix theory realization of this full
U-duality symmetry group of M-theory is, of course, closely related to 
the issue of rotational (Lorentz) invariance of matrix theory, in 
particular to the ability to change the value of the longitudinal or 
light-cone momentum. While we will not be able to resolve this
issue, we indeed find a new duality symmetry in $E_{d+1}$, associated
with the $(d+1)$'th node of the Dynkin diagram of $E_{d+1}$,
exchanging the rank (light-cone momentum) $N$
of the gauge group with quanta of flux. Both for this reason and
because this duality is reminiscent of Nahm duality
\cite{Nahm}, we will refer to this as N-duality. The necessity
to discuss all the light-cone sectors simultaneously in a more
complete formulation of the theory had been anticipated by Susskind
\cite{susskind}. A realization of N-duality symmetry in such a theory
would be an indication of its Lorentz invariance.  Nahm duality has 
recently been discussed in a closely related context 
(SYM in $(3+1)$ dimensions) in \cite{Verlinde}. 

In trying to interpret the action of this extended U-duality group
as an action on the BPS states of a SYM theory in, say, $(3+1)$
dimensions, we encounter the problem that a state with non-zero $N$
can be mapped to a state with $N=0$. Clearly this is a rather 
singular state from a SYM point of view and indeed we find that
the masses of such states diverge in the matrix theory limit.
In $(4+1)$ dimensions, such states correspond to systems of two-branes
and zero-branes on a four-torus $T^{4}$  (the absence of wrapped
four-branes implying $N=0$), and such configurations can be interpreted
by generalizing the notion of a vector bundle to that of a sheaf
(see the discussion in \cite[section 5.3]{jhgm}). 

In the case $d=3$, on the other hand, and in most other cases,
such an interpretation is not available and some new idea appears
to be required. While we were 
in the final stages of writing this paper and trying to come to terms with
this problem, papers by Connes, Douglas, and Schwarz \cite{cds}
and Douglas and Hull \cite{mdch} appeared which (if we understand them
correctly) seem to address this issue. In these papers it is argued that
in order to see the full U-duality group one needs to consider 
SYM theory on a non-commutative torus (see the references in
\cite{cds,mdch}). Indeed, \cite[eq.\ (5.14)]{cds} shows that the 
non-commutativity of the torus `regularizes' the infinite masses 
arising for $N=0$ in the commutative case. We will come back to 
this issue in section 4.2 of this paper.

This paper is organized as follows.
In section 2 we set up the framework for the discussion of
$E_{d+1}$-duality in matrix theory. We first describe the emergence 
of (the Weyl group of) $E_{d+1}$ on the M-theory side by an 
algebraic construction inspired by \cite{egkr} and then describe
the action of this U-duality group on the T-dual \IIt\ side.
In particular we describe the N-duality transformation
in terms of its action on the parameters of the \IIt\ theory
and identify it as the transformation $T^{d-1}S_{IIB}T^{d-1}$
where $T^{d-1}$ denotes a T-duality on $(d-1)$ of the circles
of the $d$-torus and $S_{IIB}$ is the S-duality of the type IIB
string theory.  
We then briefly review the matrix theory limit of the \IIt\ theory,
as described in \cite{sennew,seibergnew} as well as the BPS mass
formula for threshold and non-threshold bound-states.

In section 3, we generalize (and modify in some respects) the
algebraic analysis of \cite{egkr}. We realize the duality symmetries
described in section 2.2 as Weyl reflections in the root space of
$E_{d+1}$, determine the fundamental weights of $E_{d+1}$ in terms
of the \IIt\ parameters and indentify the highest weights unifying
the momentum and flux $E_{d}$-multiplets into a single
$E_{d+1}$-multiplet. In particular, for $d=8$ we determine the
$E_{9}$-representation that arises to be the unique integrable
representation of affine $E_{8}$ at level one, and for $d=9$ 
we find that the representation of the hyperbolic Lie algebra
$E_{10}$ is the one associated with the null-root of $E_{9}$. 

In section 4 we discuss various consequences and applications of this
extended U-duality group. We exhibit more explicitly
the action of N-duality as a Nahm duality on the quantum numbers (and
thus BPS states) of the matrix theory and we point out the difficulties
with the $N=0$ states mentioned above. We also discuss the (less
problematic as mass preserving) active interpreation of U-duality
as an action providing different pictures of the same state in
different string theories. In particular, we point out that N-duality
always provides an effectively (1+1)-dimensional SYM description 
of any bound-state involving background D$d$-branes (or D0-branes
on the M-theory/IIA side) and that this reproduces the matrix string
dictionary of \cite{dvv}. We will also argue that in even
dimensions the matrix string picture is perhaps better thought of as an
M-theory membrane theory, and this appears to be well defined even
for $d>5$. 

In the algebraic analysis of section 3, an important role is played
by a U-duality invariant $I$ for which we have a compact and unified
expression for any $d$. In section 4.5 we will show explicitly
that for $d=5,6$ this
reduces to the cubic invariant of $E_{6}$ and quartic invariant
of $E_{7}$ respectively which appear in the analysis of black
hole entropies in five and four dimensions.

Finally, we return to the $E_{d+1}$ multiplets of single-particle
states determined in section 3. In $d\leq 5$ only one new
state (the highest weight state, an $E_{d}$-singlet) is required to unify
the momentum and flux states. For $d>5$, however,
we find other new states in the Weyl group orbit of $E_{d+1}$,
one for $d=6$, $57$ for $d=7$, etc. These have no obvious 
SYM interpretation and should be indicative of the new physics
associated with lower-dimensional compactifications of string theory.
We discuss some of the (old and new) states in $d=6,7$ and identify 
concretely some of the solitonic objects responsible for the `strange' states
($\sim 1/g_{s}^{3}$) found in \cite{egkr}. In particular, we
find that - somewhat unexpectedly - the periodic Taub-NUT soliton
encountered in the context of corrections to the hypermultiplet
moduli space in the vicinity of the conifold singularity \cite{ovnsss}
makes an appearance here.

In a recent paper \cite{ch}, Hull has announced that
he has also obtained the generalization of \cite{egkr} from $E_{d}$
to $E_{d+1}$ \cite{ch2}. 

A final remark on terminology and a {\em caveat}: 
Even though, strictly speaking, for $d>3$ SYM theory itself
is only an adequate description of the matrix theory limit
of a given toroidal compactification at long wavelengths, we will
find it convenient to use a SYM-like terminology when e.g.\
discussing bound-states and the action of the U-duality group
on the quantum numbers. We will therefore also occasionally refer to 
the matrix theory limit as a `SYM' or SYM-like theory. 

The {\em caveat} regards the fact that matrix theories
are known only for $d\leq 5$ \cite{seiberg45,seibergnew,sennew,tbreview}. 
Already for $d=6$ there are problems, and for $d \geq 7$
the situation gets worse as the background configurations of
such branes are severely restricted. Nevertheless, provided that
matrix theories for $d>5$ exist and reduce at low energies to SYM, 
the considerations of this paper regarding U-duality orbits of 
BPS states should, in the same spirit as in \cite{ch}, be valid and 
provide strong constraints on the matrix theories themselves.

\section{U-Duality and Matrix Theory}

\subsection{M-Theory Facts}

Upon compactification of M-theory on a rectangular $(d+1)$-torus
$T^{d+1}$, the 
parameters of M-theory are $\ell_P$, $R_i,i = 1\dots d$, and $R_{11}$. When 
$R_{11} <\!< R_i$ this is simply the type IIA string on $T^d$. The IIA 
string theory is mapped to itself under T-duality on two different 
circles ($x_i, x_j$), and the new IIA theory will in general
have a different string coupling $g_s$.
Using the relationship between M-theory and IIA variables, 
$R_{11} = l_s g_s$ and $\ell_P^3 = \ls^3 g_s$, the two T-dualities
may be rewritten in M-theory as,
\be
R_a \rarw R_a v\;\;,\;\;\;\;\;\; \ell_P^3 \rarw \ell_P^3 v
\ee
where $a = i,j,11$ and $v = l_p^3 / R_iR_jR_{11}$. 
As we are free to choose upon which circle we compactify to reach the 
IIA string theory, this transformation has the general form 
\be
R_i \rarw {\ell_P^3\over R_jR_k}, R_j \rarw {\ell_P^3\over R_kR_i}, 
R_k \rarw {\ell_P^3\over R_iR_j}, 
\ell_P^3 \rarw {\ell_P^6\over R_iR_jR_k},
\label{ms}
\ee
where $i,j,k\in\{1\dots d+1\}$.
In \cite{egkr} this duality was interpreted 
in the context of the SYM of Matrix theory as a generalization 
of the $d=3$ S-duality. In fact, it is easy to check that 
(\ref{ms}) is precisely the transformation $T_{ijk}S_{IIB}T_{ijk}$,
where $T_{ijk}$ denotes a T-duality on the circles $i,j,k$.

The Weyl group of $E_d$ is now generated by reflections 
corresponding to this duality transformation plus 
reflections corresponding to permutation of the labelling 
of the circles $x_1, \dots x_d$. 
To extend the Weyl group from that of $E_d$ to $E_{d+1}$ 
we simply add the reflection that corresponds to 
the interchange of $x_d$ and $x_{11}$. It will be shown later that
this additional transformation has an 
interpretation in the SYM framework involving an exchange of 
the rank of the gauge group with other quantum numbers.

Our parameter space for M-theory on $T^{d+1}$ consists 
of the radii, $R_1,\dots R_d,$ $R_{11}$ and the Planck 
length $\ell_P$, a total of $d+2$ parameters. On the other hand, 
the root 
lattice of $E_{d+1}$ is $d+1$-dimensional meaning that the roots
will span a hyperplane of codimension 1. The direction not 
acted upon by the Weyl group corresponds to the quantity,
\be
I = {\ell_P^9 \over R_{11} \prod_{i=1}^d R_i} \equiv 
\frac{\ell_{P}^{9}}{R_{11}V_{R}}\label{minv}
\ee
which is an invariant under all of the above transformations.

In order to elucidate the algebraic structure of these transformations
let us consider a $(d+2)$-dimensional vector space, ${\cal M}$ with
with the standard orthonormal basis $\{m_a, a=0,\dots,d+1\}$ and 
metric $(-1,1,\dots,1)$. 
The unit vector in each direction corresponds respectively to 
$(\log\ell_P^{3},\log R_i,\log R_{11})$. Note that the string length
$\log\ls^{2}=m_{0}-m_{d+1}$ is null with respect to this metric.

The Weyl group is generated by the transformations,
\bea
R_i &\lrarw& R_{i+1}\;\;\;\;\;\;\mbox{for}\;\;i=1,\dots,d-1\non
R_d &\lrarw& R_{11}\non
R_\alpha &\rarw& R_\alpha v \;\;\;\;\;\;\mbox{for}\;\; \alpha = 0,1,2,3
\label{Weyl}
\eea
where $R_0 = \ell_P^3$ and $v=\ell_{P}^{3}/R_{1}R_{2}R_{3}$. 
In ${\cal M}$ these transformations are reflections in the vectors
\bea
\alpha_0 &=& m_0 - m_1 - m_2 - m_3\non
\alpha_i &=& m_i - m_{i+1} \;\;\;\;\;\;\mbox{for}\;\;i = 1,\ldots,d\;\;, 
\eea
acting on a vector $m \in {\cal M}$ in the standard way as
\be
m \rarw m - 2 {(m.\alpha_{a})\over \alpha_{a}^2} \alpha_{a}=m-(m.\a_{a})\a_{a}
\;\;.
\ee
Using the above metric it is easy to see that these vectors 
are the simple roots of $E_{d+1}$  and thus the reflections 
generate the Weyl group of $E_{d+1}$.

The invariant $I$ is represented in ${\cal M}$ as the vector 
${\cal I} = 3m_0 - \sum_1^{d+1}m_i$.
${\cal I}$ is orthogonal to all of the roots as it should be,
and it has norm $(d-8)$. ${\cal I}$ is of some interest to us in 
our construction as the momentum of a BPS state,
when represented by a vector in ${\cal M}$, will not generically 
be in the subspace spanned by the root lattice but will contain 
a component in the orthogonal ${\cal I}$ direction. We thus need
to project the vector onto the root lattice subspace. For a vector $m$
this projection is
\be
m \rarw m - {(m.{\cal I})\over {\cal I}^2} {\cal I}.
\ee
In terms of the momenta of the BPS states, this projection 
corresponds to a multiplication by a power of the invariant $I$,
and as this power is fixed within a given multiplet the algebraic 
structure is not modified by this additional factor.

For $d=8$, ${\cal I}$ is null and turns out to be a linear
combination of the roots of $E_9$. Thus there is no need for 
a projection onto the hyperplane orthogonal to ${\cal I}$, which is 
just as well as our projection formula is singular precisely in 
$d=8$. 

In this framework we find that all the states discussed in 
\cite{egkr} (i.e.\ the momentum and flux multiplets of $E_d$) 
are in the same 
$E_{d+1}$ multiplet and may all be generated
by the above reflections from the BPS state corresponding to 
KK-momentum in the $x_{11}$ direction, with momentum $1/R_{11}$. For instance
the reflection $\alpha_d$ takes this state to the state with 
momentum $1/R_d$ (the first state in the flux multiplet of \cite{egkr}).
$\alpha_0$ then takes this state to $R_iR_j/{\ell_P^3}$ and another 
reflection in $\alpha_d$ takes us to $R_iR_{11}/{\ell_P^3}$ (the
first state in the momentum multiplet).

In ${\cal M}$, this fundamental state is represented by the vector $-m_{d+1}$ 
and after projection onto the hyperplane spanned by the roots
we find (for $d\neq 8$) the weight vector
\be
\lambda_d = -m_{d+1} - {{\cal I}\over (d-8)}\;\;.
\ee
In fact, for all $d$, $\lambda_d$ is the fundamental 
weight of $E_{d+1}$ dual to the root $\alpha_d$. In particular,
when $d=9$ (i.e.\ for $E_{10}$) it turns out that
$\lambda_d$ is minus the
null root of the $E_9$ subalgebra of $E_{10}$. All of this 
will be described in more detail in section 3 in terms of T-dual variables
more closely related to the SYM matrix model picture to which we
now turn.

\subsection{The T-dual SYM-like Picture}

To pass from M/IIA-theory to the SYM-like variables of the dual
m(atrix) theory,
one first performs a T-duality on all the circles of the $d$-torus $T^{d}$
to arrive at what we will refer to as the \IIt\ theory. The parameters
in this theory are the dual string coupling constant $\widetilde{g_{s}}$
and the (original and dual) string length $\ls$. As the matrix theory
limit is a `double scaling limit' of $\widetilde{g_{s}}$ and $\ls$
(see section 2.3), we
will find it convenient to express everything in terms of the 
lengths $s_{i}$ of the dual $d$-torus, the coupling
constant $g^2$ of the underlying low-energy SYM theory, and the string 
length $\ell_{s}^2$ of the original 
IIA theory. $\ls$ serves to keep track of the $R_{11}=\ls  g_{s}$ of
M-theory. Various aspects of the matrix theory limit of the 
\IIt\ U-duality group will be discussed in section 4.

The basic dictionary relating the M-theory variables to those of 
the \IIt\ theory is
\bea
s_{i}&=&\frac{\ls^{2}}{R_{i}}\non
g^{2}&=&\widetilde{g_{s}}\ls^{d-3}=\frac{g_{s}V_{s}}{\ls^{3}}\;\;,
\label{dict}
\eea
where $\widetilde{g_{s}}$ is the T-dual string coupling constant
\be
\widetilde{g_{s}} = \frac{g_{s}\ls^{d}}{V_{R}}=\frac{g_{s}V_{s}}{\ls^{d}}
\ee
and $V_{s}= \prod_{i=1}^{d}s_{i}$ is the volume of the dual torus.
The Yang-Mills coupling constant $g^{2}$ follows from matching
the low-energy effective action on the D$d$-brane,
\bea
S&=&\frac{1}{\gst\ls^{d+1}}\int (\alpha')^{2}F^{2} + \ldots\non
&=&\frac{1}{\gst\ls^{d-3}}\int F^{2} + \ldots\;\;,
\eea
with the standard SYM action $\sim 1/g^{2}$.
In terms of these variables, the invariant (\ref{minv}) becomes
\be
I = \frac{g^{4}}{\vs\ls^{2(d-7)}}\;\;.
\label{yinv}
\ee
Note that this differs from the invariant considered in \cite{egkr},
\be
I_{EGKR}=\frac{V_{s}^{d-5}}{g^{2(d-3)}}\;\;,
\ee 
essentially by its dependence
on $R_{11}$, chosen in such a way as to make  $I$ $E_{d+1}$- 
and not only $E_{d}$-invariant. It turns out (see section 4.5) that
$I$ is precisely the U-duality invariant that plays an important
role in the discussion of black-hole entropies in four and five
dimensions.

In order to describe the U-duality symmetries on the \IIt\ side,
let us introduce the quantities $\gamma_{S}$ and $\gamma_{N}$
defined by
\bea
\gamma_{S}&=&\frac{g^{2}}{\prod_{i=4}^{d}s_{i}} \equiv \frac{g^{2}}{W}\non
\gamma_{N}&=&\frac{R_{11}}{R_{d}}= \frac{g^{2}\ls^{2} s_{d}}{V_{s}} 
\;\;.
\label{gammas}
\eea
$\gamma_{S}$ is the effective coupling constant in $d$ dimensions
(extrapolated form $d=3$), denoted by $g_{eff}^{2}$ in \cite{egkr}, 
and sets the scale for the S-duality transformation. $\gamma_{N}$, on
the other hand, is the effective $d$-dimensional coupling constant
extrapolated from $d=1$ and sets the scale for the N-duality
transformation $R_{d}\lra R_{11}$ in the \IIt\ picture.

On the \IIt\ side, the Weyl group of the U-duality group is generated by
the following three types of transformations (the first two of which have
been considered in \cite{egkr}): 
\begin{description} 
\item[Permutations]

This is what remains of the geometric $SL(d,\ZZ)$-duality group on a 
rectangular torus. It acts as 
\bea
g^{2}&\ra& g^{2}\non
s_{i}&\lra& s_{i+1}\non
s_{j}&\ra& s_{j} \;\;\;\;\;\;\mbox{for}\;\;j\neq i,i+1\non
\ls^{2}&\ra&\ls^{2}\;\;.
\label{yt} 
\eea

\item[S-duality]
This is the $(d+1)$-dimensional generalization of the familiar S-duality
transformation of $(3+1)$-dimensional $N\!=\!4$ SYM theory 
\bea 
g^{2}&\ra& g^{2}\gamma_{S}^{d-5}\non 
s_{i} &\ra& s_{i} \;\;\;\;\;\;\mbox{for}\;\;i=1,2,3\non 
s_{i} &\ra& s_{i} \gamma_{S} \;\;\;\;\;\;\mbox{for}\;\;i=4,\ldots,d\non 
\ls^{2}&\ra& \ls^{2} \gamma_{S}\label{ys}
\eea 
(equivalently, one can replace the first relation by 
$g_{eff}^{2} \ra 1/g_{eff}^{2}$). (\ref{ys}) is
most succinctly throught of as the transformation 
$T_{4\ldots d}S_{IIB}T_{4\dots d}$ of the \IIt\ theory, or as 
$T_{123}S_{IIB}T_{123}$ in the original IIA theory, where
$S_{IIB}$ is the S-duality transformation of type IIB string
theory and $T_{abc\ldots}$ denotes a T-duality transformation on the circles
$a,b,c,\dots$. Alternatively, (\ref{ys}) follows from applying the dictionary
(\ref{dict}) to the transformation (\ref{ms}) (with $i,j,k =1,2,3$), 

\item[N-duality]

This is the counterpart of the $R_{d}\lra R_{11}$ transformation in M-theory
and, as we will show, extends the U-duality group of SYM from
$E_{d}$ to $E_{d+1}$.
The corresponding action on the SYM-like \IIt\ 
variables is somewhat more complicated
and explicitly given by 
\bea
g^{2}&\ra& g^{2} \gamma_{N}^{d-4}\non
s_{i}&\ra&s_{i}\gamma_{N}\;\;\;\;\;\;\mbox{for}\;\;i\neq d\non
s_{d}&\ra& s_{d}\non
\ls^{2}&\ra& \ls^{2}\gamma_{N}\label{yn}
\eea
\end{description}
and the action on the \IIt\ string couping constant
$\gst$ is 
\be
\gst\ra\gst\gamma_{N}^{\frac{d-5}{2}}\;\;.
\label{yngs}
\ee
It is readily checked explicitly that $I$ (\ref{yinv}) is
invariant under all of the above transformations. 
It can also be checked that this N-duality transformation 
relates the states in the momentum and flux multiplets 
to each other (cf.\ section 2.4), as it should from the M-theory
point of view. We need to carefully distinguish between the 
manner in which the U-duality of Matrix theory on $T^d$ 
arises here and in recent works on the original matrix theory 
limit of \cite{bfss}. In our case, the transformations really
arise in the \IIt\ theory as an action on the entire collection of 
$U(N)$ SYM-theories (for all $N$) in $d+1$ dimensions, and not 
for instance on the spectrum of a fixed $U(N)$ SYM theory in 
$(d+1)+1$ dimensions as in \cite{rozali,bcd}. This is a consequence
of working in the DLCQ of M-theory.

In general, there are two ways of looking at the above transformations,
as passive transformations, in which one interprets them as acting on 
BPS-states, or the lattice of quantum numbers, and as active transformations
on the parameters $(g_{s},\ls,s_{i})$ of the string theory. 
Passively, S-duality exchanges for example the electric and magnetic quantum 
numbers $(e_{i},m_{jk})$ in the directions $1,2,3$, 
\be
\mbox{Passive S-duality:}\;\;e_{1}\lra m_{23}\;\;\mbox{etc.}
\ee
Thought of actively, S-duality acts, as mentioned above, as 
\be
\mbox{Active S-duality:}\;\;T^{d-3}S_{IIB}T^{d-3}\;\;.
\ee

The most interesting aspect of N-duality is that,
as we will explain in detail in section 4, in the passive interpretation
it acts non-trivially on the
rank $N$ of the gauge group $U(N)$ of the underlying low-energy SYM theory,
exchanging it with the electric flux quantum number $e_{d}$ in the 
$d$-direction, 
\be
\mbox{Passive N-duality:}\;\;e_{d}\lra N\;\;\mbox{etc.}
\ee
Thus, N-duality is reminiscent of Nahm-duality \cite{Nahm}. This symmetry 
has previously been discussed in the context of the U-duality symmetry of
$(3+1)$-dimensional SYM theory in \cite{Verlinde}. Interpreted
actively, on the other hand, N-duality acts as
\be
\mbox{Active N-duality I:}\;\;T^{d-1}S_{IIB}T^{d-1}\;\;,\label{anI}
\ee
where $T^{d-1}$ is shorthand for the $(d-1)$-fold T-duality on the
circles transverse to $s_{d}$.
This can be seen either from the M-theory origin of the IIB S-duality
$S_{IIB}$, or directly in terms of the action on the parameters
$(s_{i},s_{d},\gst,\ls^{2})$:
\bea
(s_i,s_d,\gst,\ls^{2}) 
&\stackrel{T^{d-1}}{\longrightarrow}& 
(\frac{\ls^{2}}{s_i},s_d,\gst\frac{\ls^{d-1}s_{d}}{V_{s}},\ls^{2}) \non
&\stackrel{S_{IIB}}{\longrightarrow}& 
(\frac{\ls^{2}}{s_i},s_d,\frac{V_{s}}{\gst\ls^{d-1}s_{d}},
\frac{\ls^{d+1}\gst s_{d}}{V_{s}}) \non
&\stackrel{T^{d-1}}{\longrightarrow}& 
(s_i \gamma_{N},s_d,\gst\gamma_{N}^{\frac{d-5}{2}},\ls^{2}\gamma_{N}) 
\label{activeaction}
\eea
In particular, in the matrix theory limit (cf.\ section 2.3), 
the N-dual theory becomes effectively $(1+1)$-dimensional,
\be
\mbox{Active N-duality II:}\;\;\mbox{SYM}_{d+1}\ra \mbox{SYM}_{1+1}\;\;.
\label{anII}
\ee
precisely as in \cite{dvv}. We will say more about
these active and passive interpretations of N-duality in section 4.

What is probably not at all obvious at this point is that this N-duality
really extends the U-duality group from $E_{d}$ to $E_{d+1}$. This is
most easily verified within the appropriate generalization of the 
algebraic framework introduced in \cite{egkr} which we will construct 
in section 3. 

\subsection{The Matrix Theory Limit and the DLCQ of M-Theory}

Shortly after the original BFSS m(atrix) theory proposal \cite{bfss}
for a non-perturbative definition of M-theory, it was realized that
toroidally compactified string theory is related to (the large $N$
limit of) maximally supersymmetric $U(N)$ Yang-Mills theory on the 
dual torus \cite{taylor,bfss} (this prescription has recently been 
reanalyzed in \cite{cds}, with some rather striking consequences). 
In fact, this picture follows naturally from T-duality, 
mapping a configuration of $N$ D0-branes on the IIA side to that of 
$N$ D$d$-branes on the dual \IIt\ (= IIA or IIB) side. In this setting,
the SYM theory arises as the low-energy effective action on the 
world-volume of the D$d$-branes. For several reasons, however, this 
picture is incomplete and not completely satisfactory. 

First of all, for $d>3$ the SYM theory is not
renormalizable and thus gives an incomplete description of the physics
at short distances. This (among other things) led to the search for more
general (field, string, \ldots?) theories on the world-volumes of extended
objects in string theory or M-theory which are decoupled from the ten-
or eleven-dimensional bulk dynamics. Subsequently, a matrix theory description 
for $d=4,5$ was proposed in terms of a $(2,0)$ supersymmetric field theory on 
the M5-brane \cite{rozali,seiberg45} and a theory of `microstrings' localized 
on the world-volume of the NS5-brane \cite{dvvold,seiberg45,dvv} 
respectively. 

Secondly, many of the attractive features of the original BFSS theory
(and its descendants) appear strictly speaking only in the infinite
momentum frame $N\ra\infty$ limit which is awkward to deal with directly.
Therefore the suggestion by Susskind \cite{susskind} that the finite
$N$ matrix model might actually give a complete description of the discrete
light cone quantization (DLCQ) of M-theory in the sector carrying $N$  units
of light-like momentum, was particularly attractive. 

Very recently, Sen and Seiberg \cite{sennew,seibergnew} combined these
two issues to provide a) substantial evidence for Susskind's conjecture 
and b) a systematic way of deriving the matrix models proposed 
for $d\leq 5$. In particular, therefore, this means that one can 
consider the `double scaling' matrix theory limit $\ls\ra 0, 
g_{s}\ra 0$ of M-theory compactified on space-like circles 
(a setting in which the $E_{d+1}$ U-duality is manifest on 
the M-theory side) and reach the finite $N$ 
DLCQ of M-theory in the limit, assigning a physically meaningful 
interpretation to the finite $N$ versions of the matrix models proposed 
for $d\leq 5$. One might perhaps have expected the U-duality group
of a light-like torus to differ from that of a space-like torus.
However, this is apparently not the case \cite{ch,hulljulia}.

In terms of the M/IIA-theory variables, this matrix theory limit is
the limit $\ls\ra 0,g_{s}\ra 0$, keeping fixed the radii $R_{i}$ 
measured in Planck units $\ell_{P}$, and the ratio $\ls^{3}/g_{s}$.
In particular, therefore, the ratios $R_{i}/\ls^{2}$ are fixed (thus
$R_{i}\ra 0$) and the matrix theory limit is actually
\bea
\mbox{IIA Matrix Theory Limit:}&& \ls\ra 0\;\;,\;\;g_{s}\ra 0\non
                             && R_{i}/\ell_{P}=\mbox{constant} \non
                             && R_{i}/\ls^{2}=\mbox{constant}\;\;.
\eea
The more transparent interpretation of this on the \IIt\ side is 
that one is taking the same limit, keeping fixed the parameters
(\ref{dict}), i.e.
\bea
\mbox{\IIt\ Matrix Theory Limit:}&& \ls\ra 0\;\;,\;\;g_{s}\ra 0\non
                             && s_{i}=\mbox{constant} \non
                             && g^{2}=\mbox{constant}\;\;.
\eea
This limit captures correctly the SYM degrees of freedom and is also
natural from the D-brane probe point of view \cite{maldacena}.

{}From (\ref{dict}) we can read off that in $d\leq 2$ the matrix theory 
limit amounts to letting $\ls\ra 0$ and $\gst\ra 0$, while in $d=3$
one has $\ls\ra 0$ with $\gst$ constant. In both cases, this amounts to
decoupling of the dynamics on the branes from the bulk dynamics (in 
particular, gravity). In these cases, thus, the SYM picture provides an
adequate description of the DLCQ of M-theory. 

In $d=4$, $\gst\ra\infty$ in the matrix theory limit, and thus 
one can certainly not ignore the coupling to the bulk fields.
However, as $\gst\ls$ is constant in the matrix theory limit, this
suggests the emergence of a new `eleventh' direction in the \IIt\ theory
with radius $\widetilde{R}_{11}=\ls\gst$, whose $\widetilde{\mbox{D}0}$
branes are the instantons on the D4-brane worldvolume \cite{rozali}. 
The dynamics on the five-brane (the $(2,0)$ supersymmetric theory) of this 
$\widetilde{M}$-theory also decouples from the bulk dynamics because
$\widetilde{\ell_{P}}^{3}=\ls\gst^{3}\ra 0$ \cite{seiberg45}. 

Also in $d=5$, $\gst$ diverges in the matrix theory limit. In this
case, one can perform an S-duality to convert the $N$ D5-branes to
$N$ solitonic NS5-branes. The new string coupling
constant $\widehat{g_{s}}=1/\gst \ra 0$, while the new string length
$\widehat{\ls}^{2}=g^{2}$ remains constant, suggesting that at these
energies the theory is described by string-excitations confined to
the NS5-brane worldvolume \cite{dvvold,seiberg45,dvv}. 

In $d>5$, in spite of a number of attempts \cite{d6}, no satisfactory 
matrix theory formulation is known \cite{seibergnew,sennew}. However,
one expects that, if found, this theory will reduce at low energies
to SYM theory in $(d+1)$-dimensions, and that (part of) the BPS spectrum
can be reliably determined at low energies. With this in mind, 
we will make use of the matrix theory limit in section 4 to extract 
information about the U-duality groups of matrix theories from those of 
the \IIt\ theories.

\subsection{D$d$-Brane Backgrounds and Bound States}

As we will frequently make use of the BPS mass formulae in the matrix 
theory limit in subsequent sections, we briefly recall the relevant
equations.

Let us consider $N$ D0-branes on the M/IIA-side.
Upon passing to the \IIt\ side, this gives a configuration of
$N$ D$d$-branes wrapping the (dual) torus $T^{d}$. This state
has a mass
\be
M_{Nd}=\frac{N}{R_{11}}=\frac{NV_{s}}{\gst\ls^{d+1}}\;\;.
\label{mndd}
\ee
In order to analyze the energy of BPS bound states in the matrix theory 
(SYM) limit, it will be convenient to reexpress (\ref{mndd}) in terms of 
the \IIt\ variables $(g^{2},s_{i},\ls^{2})$ as 
\be
M_{Nd}=\frac{NV_{s}}{g^{2}\ls^{4}}\;\;.
\label{nd}
\ee
In the matrix theory limit, this state becomes infinitely heavy and
should thus be treated as a background configuration whose energy
is to be subtracted in the calculation of bound state energies.

BPS bound states of the background D$d$-branes with other objects
in the theory (KK modes, wound strings, D-branes, solitons) can basically
be of two kinds. There are bound states which preserve half of the original
supersymmetries of the IIA theory, i.e.\ no further supersymmetries
are broken beyond those broken by the background. These are non-threshold
bound states with non-zero binding energy and, by the saturation of the
Bogomoln'y bound for the BPS state, the energy $E$ of the 
bound state is related to the background mass (\ref{nd}) and the mass 
$M$ of the other object by
\be
E^{2}=M_{Nd}^{2}+M^{2}
\ee
Examples of such bound states are D0-KK and D0-D2 systems on the IIA 
side, and thus bound states of the background D$d$-branes with either
wound NS strings or wrapped D$(d-2)$-branes on the \IIt\ side. An NS
string wound $p$ times around the circle $s_{i}$ has a mass
$M=ps_{i}/\ls^{2}$, and thus the bound state energy is
\be
E^{2}=\frac{N^{2}V_{s}^{2}}{g^{4}\ls^{8}}+\frac{p^{2}s_{i}^{2}}{\ls^{4}}
\;\;.
\ee
Calculating the energy in the matrix theory limit $\ls\ra 0$ and subtracting
the background one finds
\bea
E_{SYM}&=& \lim_{\ls\ra 0}\left(
\sqrt{\frac{N^{2}V_{s}^{2}}{g^{4}\ls^{8}}+\frac{p^{2}s_{i}^{2}}{\ls^{4}}}
- \frac{NV_{s}}{g^{2}\ls^{4}}\right)\non
&=& \frac{p^{2}g^{2}s_{i}^{2}}{2NV_{s}}\;\;, 
\eea
i.e.\ precisely the energy of an electrix flux state of SYM theory. More
generally, whenever $M\sim\ls^{-2} + O(\ls^{-1})$ as $\ls\ra 0$, 
one will obtain a finite (and non-zero)
energy 
\be
E_{SYM}=\frac{M^{2}}{2\mnd}
\ee
in the matrix theory limit (we will omit this factor of $1/2$ in the
following). In particular this is also the case for the
p-wrapped D$(d-2)$-brane with mass ($i,j$ are the unwrapped directions)
\be
M=\frac{pV_{s}}{s_{i}s_{j}\gst\ls^{d-1}}=
\frac{pV_{s}}{s_{i}s_{j}g^{2}\ls^{2}}\;\;,
\ee
leading to the standard magnetic flux bound state energy
\be
E_{SYM}= \frac{p^{2}V_{s}}{Ng^{2}s_{i}^{2}s_{j}^{2}}\;\;.
\ee
in the limit.

The other class of bound states of interest to us are so-called 
threshold bound states preserving just one quarter of the original
supersymmetries. These are subtle objects in general, but the 
prototypical examples here are D0-NS winding
and D0-D4 states on the IIA side, corresponding to momentum states
$M=p/s_{i}$ and D$d$-D$(d-4)$ bound states on the \IIt\ side. As the
binding energy is zero, these satisfy the linear bound state energy
relation
\be
E=\mnd+M\;\;,
\ee
so that 
\be
E_{SYM}=\lim_{\ls\ra 0}M\;\;.
\ee
This will be finite and 
non-zero in the limit if $M \sim \mbox{constant} + O(\ls)$ as $\ls\ra 0$. 
This is the case, in particular, for the above momentum state, but also for
the $p$-wrapped D$(d-4)$-brane which has mass
\be
M=\frac{pV_{s}}{\gst\ls^{d-3}s_{i_1}s_{i_2}s_{i_3}s_{i_4}}
=\frac{pV_{s}}{g^2 s_{i_1}s_{i_2}s_{i_3}s_{i_4}}
\ee
and represents an instanton in the matrix theory. We will encouter 
more exotic examples of matrix theory bound states in section 4.

It is now also easy to see in which sense N-duality connects the flux
and momentum multiplets of \cite{egkr}.
E.g.\ starting with the momentum state with $M=1/s_{i}$ one obtains 
the magnetic flux state with mass $M=V_{s}/g^{2}\ls^{2}s_{i}s_{d}$.
However, as N-duality acts non-trivially on the background, in order
to have a SYM-like interpretation in both cases, one needs
to start with a configuration with non-zero momentum in both the 11- and
the $d$-direction. This will also be discussed in more detail in section 4,
though already now it is clear that all the one-particle states in the
flux and momentum multiplets can be generated from the background (vacuum)
state $\N=1/R_{11}$ (as noted in section 2.1) or
\be
\N=\frac{V_{s}}{g^{2}\ls^{4}}
\label{caln}
\ee
by U-dualities. In fact (cf.\ section 3.4), $\N$ is precisely the highest 
weight state in the orbit of the U-duality group, and that the highest
weight state corresponds to the vacuum does probably not come as a surprise.

At various points in this paper we talk about BPS field configurations in
terms of their (gravitational) 
masses. For $d\leq 6$ this is perfectly legitimate but
for $d>6$ the concept of mass in the $(9-d)+1$ dimensional non-compact
spacetime is ill-defined due to the lack of asymptotic flatness of the
configurations. Nevertheless, also for $d>6$ $E_{SYM}$ makes sense and
is the appropriate quantity to consider for the purposes of (formally)
discussing the action of U-duality on the BPS states.

\section{Extending the U-Duality Group from $E_{d}$ to $E_{d+1}$: The Algebra}

\subsection{The Algebraic Set-Up}

In \cite{egkr}, a convenient framework for discussing various 
algebraic aspects of U-duality was introduced. We will make
use of this framework here, with some modifications which make
the construction more natural from an M-theory/$E_{d+1}$ point of view.

Thus consider a $(d+2)$-dimensional vector space $\cal Y$ spanned by 
$(\log g^{2}, \log s_{i}, \log \ls^{2})$ and introduce the \IIt\ 
counterparts $\{y_{a}\}$ 
of the M-theory vectors $\{m_{a}\}=\{\log \ell_{P}^{3},\log R_{i},
\log R_{11}\}$, namely
\bea
y_{0}&=& \log g^{2} + 3\log \ls^{2} -\lv \non
y_{i}&=& \log\ls^{2} -\log s_{i}\non
y_{d+1}&=& \log g^{2} + 2\log \ls^{2} -\lv\equiv -\log\N\;\;.
\eea
One can now postulate (or deduce from the flat M-theory metric) the metric
\be
y_{a}.y_{b} = \eta_{ab}\;\;\;\;\;\;\eta_{ab}=\diag(-1,+1,\ldots,+1)\;\;.
\ee
In terms of the natural \IIt\ or
matrix theory variables $(\gst,s_{i},\ls^{2})$ or $(g^{2},s_{i},\ls^{2})$ 
this becomes the off-diagonal metric $G_{ab}$ with non-zero components
\bea
G_{\gst\gst}=2
&\mbox{or}& 
G_{g^{2}g^{2}} = (5-d) 
\non
G_{\gst\ls^{2}} = -1
&\mbox{or}& 
G_{g^{2}\ls^{2}}=-1 
\non
&G_{s_{i}s_{i}}=1&\;\;.
\eea
In $\cal Y$, the invariant $I$ (\ref{yinv}) is represented by the
vector
\be
{\cal I} = 2\log g^{2} - \log V_{s} - (d-7) \log \ls^{2}\;\;,
\ee
with norm 
\be
{\cal I}.{\cal I} = (d-8)\;\;.
\ee
We thus see that for $d\leq 7$ the induced metric on the 
orthogonal complement to $\cal I$ in $\cal Y$ is positive 
definite, while for $d=9$ it is indefinite. The case
$d=8$ is special and will lead to a degenerate metric 
on root space. All this is, of course, consistent with
the structure one expects for the root space of $E_{d+1}$.

\subsection{The Weyl Group of $E_{d+1}$}

It is now possible to realize the transformations (\ref{yt}-\ref{yn})
as reflections in the vector space $\cal Y$. Indeed, these
transformations are reflections along the vectors
\bea
\a_{0} &=& \log g^{2} - \log W=\log\gamma_{S}\non
\a_{i} &=& \log s_{i+1} -\log
s_{i}\;\;\;\;\;\;\mbox{for}\;\;i=1,\ldots,d-1\non
\a_{d}&=&-\log \gamma_{N}\label{yroots}
\eea
respectively. In this setting, the invariance of $I$ is expressed by
\be
\a_{0}.{\cal I} = \a_{i}.{\cal I} = \a_{d}.{\cal I} = 0\;\;.
\ee
The following properties of the $\a_{a}$ are now readily verified:
\bea
(\a_{a})^{2} &=& 2 \;\;\;\;\;\;\forall\;\;a=0,\ldots,d\non
\a_{i}.\a_{i+1} &=&-1 \;\;\;\;\;\;\forall\;\;i=1,\ldots,d-1\non
\a_{0}.\a_{3}&=&-1\;\;.
\eea
Thus the $\{\a_{a}\}$ represent precisely the root system and 
Dynkin diagram of $E_{d+1}$ and the U-duality symmetries 
(\ref{yt}-\ref{yn}) generate the Weyl group of $E_{d+1}$.

Let us be slightly more explicit about this for the `exotic' cases
$E_{9}$ and $E_{10}$. $E_{9}=\widehat{E_{8}}$ is the affine algebra
of $E_{8}$. Its root system  is usually presented in terms of the
vectors
\bea
\bar{\a}_{i}&=&(\beta_{i},0,0)\non
\delta &=& (0,0,1)\non
\kappa &=& (0,1,0)\;\;,
\eea
where $\beta_{i}$, $i=0,\ldots,d-1=7$
are the simple roots of $E_{8}$ and the non-vanishing
scalar products are
\be
\bar{\a}_{i}.\bar{\a}_{j} = \beta_{i}.\beta_{j}\;\;\;\;\;\;
\delta.\kappa = 1\;\;.
\ee
The simple roots are chosen to be $\bar{\a}_{i}$ and 
\be
\bar{\a}_{8}=(-\psi,0,1)\;\;,
\ee
where $\psi$ is the highest root of $E_{8}$. 

To identify this structure in the present context, we can first of all
identify the $\bar{\a}_{i}$ with our $\a_{i}$ for $i=0,\ldots,7$. 
Explicit calculation shows that the remaining root $\a_{d=8}$ (\ref{yroots})
can be written as
\be
\a_{d=8} = - \psi + {\cal I}\;\;.
\ee
In particular, therefore, $\cal I$ is a linear combination of the roots,
\be
{\cal I} = 3\a_{0} + 2 \a_{1} + 4\a_{2} + 6 \a_{3} + 5\a_{4} + 4 \a_{5}
+ 3\a_{6} + 2\a_{7} + \a_{8}\;\;.
\label{i8r}
\ee
so that we can identify $\cal I$ with a multiple of $\delta$,
\be
{\cal I} = p \delta\;\;.\label{yipd}
\ee
Thus we have $\bar{\a}_{8}=-\psi + {\cal I}/p$ and $\a_{d=8} = -\psi
+ p\delta$. Finally, there is another null-direction in
${\cal Y}$, orthogonal to all the roots of $E_{8}$, spanned by
\be
{\cal K} = {\cal I} -2\log \N\;\;.
\label{yk8}
\ee
$\cal I$ and $\cal K$ have a non-zero scalar product, ${\cal I}.{\cal K}
=-2$, and we we can thus identify ${\cal K} = q\kappa$. The scalar product 
$\kappa.\delta =1$ now implies that $pq=-2$ or that
\be
{\cal K} = -\trac{2}{p}\kappa\;\;.\label{ykqk}
\ee
At this point, $p$ is undetermined. However, we will see below that
both the considerations regarding $E_{10}$ and those regarding the 
representation of the U-duality group in $d=8$ imply that $p=1$
so that
\bea
\a_{d=8} &=& \bar{\a}_{8} = (-\psi,0,1)\non
{\cal I} &=& \delta\non
{\cal K} &=& -2\kappa\;\;.
\eea

Let us now turn to the U-duality group $E_{10}$ that we obtain for $d=9$.
$E_{10}$ is a hyperbolic Lie algebra (meaning a Kac-Moody algebra which
is such that upon removal of any node from its Dynkin diagram one obtains 
either finite-dimensional or affine Kac-Moody algebras - for a digestible 
introduction to $E_{10}$ see \cite{e10}). 
Recall that we have the roots
$\{\a_{a}, a =0,\ldots, 9\}$, of which we identify the $\a_{i}$,  
$i=0,\dots,7$, with the simple roots of $E_{8}$. It is conventional to
replace $\a_{8}$ by the null root $\delta$ of $E_{9}$, i.e.\ by 
${\cal I}_{d=8}/p$, where ${\cal I}_{d=8}$ is the eight-dimensional
invariant, given in terms of roots in (\ref{i8r}). Its scalar 
product with the hyperbolic root $\a_{d=9}$ is known to be
\be
\a_{d=9}.\delta = -1\;\;,
\ee
but, as ${\cal I}_{d=8}= \a_{8} + \ldots$, one has $\a_{d=9}.{\cal I}_{d=8}
=-1$, and thus we find $p=1$. The other null direction orthogonal to
${\cal I}_{d=9}$ and the roots of $E_{8}$ is spanned by $\a_{d=9}+\delta$.
This is directly related to the $\kappa$ of $E_{9}$. In fact, while
${\cal K}_{d=8}$ (\ref{yk8}) constituted an independent direction 
in $E_{9}$,
in $E_{10}$ it can be expressed in terms of the roots as 
\be
{\cal K}_{d=8} = 2(\a_{d=9}+\delta)\;\;,
\ee
so that
\be
\kappa = -(\a_{d=9} +\delta)\;\;,
\ee
with
\be
\kappa^{2}=0\;\;,\;\;\;\;\;\;\kappa.\delta =1\;\;.
\ee
 
\subsection{Fundamental Weights of $E_{d+1}$}

In \cite{egkr}, primarily two kinds of BPS states were
investigated, the momentum multiplet, associated on the M-theory
side with longitudinally wrapped branes, and the flux multiplet,
corresponding to KK modes (electric flux) or transversally wrapped
branes (magnetic flux). These were shown to transform seperately under
the $E_{d}\subset E_{d+1}$ duality group (or rather, its Weyl group)
generated by S-duality and permutations. 

{}From the M-theory point of view, it is of course obvious that including
$R_{11}\lra R_{d}$ permutations (upon which the distinction between
longitudinal and transverse branes disappears) will connect these two
multiplets, and we have also already indicated this explicitly 
in sections 2.1 and 2.4 above. It follows that the 
two $E_{d}$-multiplets are part of one $E_{d+1}$ multiplet, and in 
the following we will identify the corresponding highest weight 
states, deferring to section 4 a discussion of the new BPS states one 
obtains in that way (states that are neither in the
momentum nor in the flux multiplet of $E_{d}\subset E_{d+1}$).

As all the $E_{d+1}$ are simply laced, the fundamental weights $\{\la_{a}\}$ 
dual to the simple roots $\{\a_{b}\}$ (\ref{yroots}) of $E_{d+1}$
are characterized by $\la_{a}.\a_{b} = \d_{ab}$.
Furthermore, looking for solutions to these equations within the 
$(d+2)$-dimensional vector space ${\cal Y}$, for $d\neq 8$ we need to
require that the $\la_{a}$ be orthogonal to the invariant ${\cal I}_{d}$.
For $d=8$, on the other hand, the fundamental weights are only determined
up to addition of a multiple of the null root $\delta$ (or the invariant
${\cal I}_{d=8}$). As the null root is orthogonal to all the roots
of $E_{9}$, this does not change the corresponding highest weight
representation. For reference purposes, we here provide explicit 
expressions for all the fundamental weights $\{\la_{a}\}$. 

For $d\neq 8$, the fundamental weights dual to the simple roots 
(\ref{yroots}) and orthogonal to ${\cal I}_{d}$ in ${\cal Y}$
are
\bea
\la_{0}&=& \trac{1}{(d-8)}[(2-d)\lg +(d-5)\lv +3\log\ls^{2}]\non
\la_{i}&=& \trac{i}{2(d-8)}[(4-d)\lg +(d-6)\lv +2\log\ls^{2}] - \log V_{s,i}
\;\;\;\;\;\;i=1,2,3\non
\la_{j}&=& \trac{1}{(d-8)}[(-3d+3j+6)\lg +(3d-j-15)\lv +(9-j)\log\ls^{2}]\non
&& -\log V_{s,j}\;\;\;\;\;\; j=4,\ldots,d-1\non
\la_{d} &=& \trac{1}{(d-8)}[(6-d)\lg +(d-7)\lv +(9-d)\log\ls^{2}]\;\;,
\eea
where $V_{s,i}$ denotes the partial volume $\prod_{a=1}^{i}s_{a}$.     

As mentioned above, the fundamental weights are not unique in
$d=8$, but a convenient set of representatives is 
\bea
\la_{0}&=& -\lg + \lv -3\log\ls^{2}\non
\la_{i}&=& i(\log V_{s}-\log g^{2}-2\log\ls^{2}) 
-\log V_{s,i}\;\;\;\;\;\;i=1,2,3\non
\la_{j}&=& -3\lg +3\lv +(j-9)\log\ls^{2}-\log V_{s,j}
\;\;\;\;\;\;j=4,5,6,7\non
\la_{d=8} &=& \log V_{s} -\log g^{2} - 2 \log\ls^{2} =\log\N\;\;.
\eea

In general, because of the necessity to project the weights, and
because the invariant $I$ is dimensionful, the correspondence 
between the masses of BPS states and weights of the U-duality 
group is somewhat indirect. Both of these complications are 
absent in $d=8$, however, and some conclusions can be drawn 
directly from looking at the highest weight states. It is easy
to see that the mass-dimensions of the highest weight states
of the fundamental representations are 
\be
{}[\la_{0}]=3, [\la_{i}]=2i, [\la_{j}]=9-j, [\la_{d=8}]=1\;\;.
\ee
Thus the only highest weight (positive integral linear combination
of the fundamental weights) having a mass dimension of an energy
is $\la_{d=8}$, and this is indeed the highest weight state of the
half-BPS configurations. It follows that
the U-duality orbits of other BPS bound states are necessarily described
in terms of higher powers of the mass. We see that this 
fact, which is well known for $E_{d\leq 8}$  (see e.g.\ \cite{fm}), 
follows rather readily for $E_{9}$ within the present framework.

\subsection{The Highest Weights of the U-Duality Group $E_{d+1}$}

The fundamental weight we will primarily be interested in is the 
weight $\la_{d}$ dual to the root $\a_{d}$ generating the N-duality
transformation (\ref{yn}). It is always of the form 
\be
\la_{d} = \ln + \mu_{d}{\cal I}_{d}
\ee
for some $d$-dependent constant $\mu_{d}$, where $\N$ was defined in
(\ref{caln}), and as such represents the precise SYM counterpart
of the M-theory longitudinal momentum state $1/R_{11}$ discussed 
in section 2.1. As explained in section 2.4, $\N$ just represents the
D$d$-brane background of the $\widetilde{\mbox{II}}$ theory, 
i.e.\ the vacuum, a not unexpected feature of a highest weight state.

Of course, as a result of the enlargement of the geometric duality
group on the M-theory side from $SL(d,\ZZ)$ to $SL(d+1,\ZZ)$, 
we also find states which appear neither in the momentum nor in
the flux multiplets of $E_{d}$. In $d\leq 5$
dimensions, only one new state, the highest weight state
$\N$, is needed to unify the momentum and flux multiplets.
But e.g.\ for $d=6$, instead of 
the $27+27=54$ flux and momentum states in the $\rb{27}\oplus\r{27}$
of $E_{6}$ \cite{egkr}, we find $56$ states in the Weyl group orbit of the
$\r{56}$ of $E_{7}$. And for $d=7$, instead of 
the $56+126=182$ flux and momentum states in the $\r{56}\oplus\r{133}$
of $E_{7}$  we find $240$ states in the Weyl group orbit of the
$\r{248}$ of $E_{8}$. We will return to this in section 4.

In $d=8$, the U-duality group is $E_{9}=\widehat{E_{8}}$. Again, the 
highest weight of interest is the fundamental weight dual to $\a_{d}$,
i.e.\ $\la_{d}=\ln$ or (\ref{yk8})
\be
\la_{d=8}=\frac{1}{2}({\cal I}-{\cal K})\;\;.
\ee
Using (\ref{yipd}) and (\ref{ykqk}), we can write this as
\be
\la_{d=8}= \frac{1}{2}(p\delta +\frac{2}{p}\kappa)
=(0,\frac{1}{p},\frac{p}{2})\;\;.
\ee
We can now fix $p$ (and hence the level of the representation) by an 
argument analogous to (but more conclusive than) that employed in
\cite{egkr} for $d=9$. First of all, we  observe that demanding that
the level be an integer imposes  the requirement $(1/p)\in\ZZ$. 
Secondly, since 
\be
\a_{d=8}=-\psi + p\delta\;\;,
\ee
the requirement that the affine algebra have an integer moding
imposes $p\in\ZZ$. These two conditions are uniquely solved
by $p=1$ (as we also found via $E_{10}$ in section 3.4). Thus
the level is $k=1$, and in fact the representation $\la_{d=8}$
is the unique integrable representation of $E_{9}$ at $k=1$. 
Why one should find a unitarizable representation (or if there
is a good reason for this) remains a mystery at this point \cite{egkr}.

Finally, for $d=9$, i.e.\ $E_{10}$, the highest weight state is
\be
\la_{d=9}=2\lv - 3\lg\;\;.
\ee
As $\la_{d=9}$ is null and orthogonal to all the simple roots
apart from $\a_{d=9}$, it is proportional to the null root $\delta$
of $E_{9}$, and $\a_{d}.\la_{d}=1$ determines 
\be
\la_{d=9}=-\delta\;\;.
\ee
While this is certainly a distinguished highest weight of $E_{10}$,
in the absence of some information about this representation of $E_{10}$
(which we have not been able to find) there is nothing that we can add 
to this at this point.

\section{Interpretation and Applications}

We have seen in the considerations of the previous sections that
N-duality, as defined in (\ref{yn}), and concretely realized as
the transformation $T^{d-1}S_{IIB}T^{d-1}$ (\ref{anI}), is the
natural transformation to add to the Weyl group of $E_{d}$ to
extend it to the Weyl group of the full U-duality group $E_{d+1}$.
In fact, N-duality is precisely the transformation associated
with the $(d+1)$'th node of the Dynkin diagram of $E_{d+1}$.

In order to clarify the general structure of the  
N-duality transformation (\ref{yn}), in this section we
will describe in more detail how it acts on the BPS quantum 
numbers, in which sense it can be regarded as a Nahm-like
duality, and in which sense it provides a $(1+1)$-dimensional 
or $(2+1)$-dimensional picture of BPS states (and perhaps not 
only of those). We also relate the U-duality invariant
$I$ (\ref{minv},\ref{yinv}) to the expression for entropies of black holes
in four and five dimensions, we look at the BPS spectra 
in $d=6$ and $d=7$ and provide concrete interpretations for some of the 
states we find.

\subsection{U-Duality, Nahm Duality, and the Matrix Theory Limit}

Let us introduce the units of electric, magnetic,
and momentum flux $\E_{i}$, $\M_{ij}$, $\P_{i}$ as well as the
quantity $\N$, defined by
\bea
\N&=& \frac{V_{s}}{g^{2}\ls^{4}}\non
\P_{i} &=&\frac{1}{s_{i}}\non
\E_{i} &=& \frac{s_{i}}{\ls^{2}}\non
\M_{ij}&=& \frac{V_{s}}{g^{2}\ls^{2}s_{i}s_{j}}\;\;\;\;\;\;i<j\;\;.
\eea
These correspond to the $(d+1)$-dimensional KK and wrapped M2-brane masses
on the M-theory side. Let us denote the corresponding quantum numbers
by $(N,p_{i},e_{i},m_{ij})$ (in general, we should of course also introduce
analogous quantum numbers for wrapped M5-brane states etc.).
S-duality in the $i,j,k$ directions
exchanges $\E_{i}$ and $\M_{jk}$, or $e_{i}$ and $m_{jk}$,
leaving all the other quanta invariant, while N-duality acts as
\bea
\E_{d}\lra \N&& \mbox{i.e.}\;\;e_{d}\lra N\non
\P_{i}\lra \M_{id}&& \mbox{i.e.}\;\;p_{i}\lra m_{id}
\label{ynq}
\eea
for $i<d$, leaving the other quantum numbers invariant.
We are not being careful with signs here but these can
be readily deduced from the T-duality rules for D-branes
given in \cite{dbtd}. For example for $d=3$ one finds that the
ten quantum numbers $(N,p_{i},e_{i},m_{i})$, corresponding
to the background D3-branes, KK modes, and fundamental and
D-string winding numbers, transform as ($a=1,2$)
\bea
N\ra -e_{3} & p_{a}\ra \epsilon_{ab}m_{b} \non
e_{3}\ra N & m_{a}\ra \epsilon_{ab}p_{b} 
\eea
with $(p_{3},m_{3},e_{a})$ invariant, much as in \cite{Verlinde}.
Arranging the ten quantum numbers as an anti-symmetric $(5\times 5)$
matrix, one can verify that this transformation is indeed implemented
by an $SL(5,\ZZ)$-rotation, as expected.

In particular, we recover the fact mentioned before that N-duality relates
states in the momentum multiplet to magnetic flux states in the flux
multiplet. 
However, the most interesting effect of this transformation
is encoded in the first line of (\ref{ynq}). As we have seen,
$\N$ is to be interpreted 
as measuring the bound state energy of the background D$d$-branes,
and therefore the corresponding quantum number is the rank of the 
gauge group. 

A duality action in SYM theory acting non-trivially on the rank
of the gauge group is not as unfamiliar as it may at first seem, at least
in the context of D-branes. Consider for example a bound state of $N$ 
D$d$-branes on $T^{d}$ (giving rise to $U(N)$ SYM theory) together with
a D$(d-2)$-brane wound $M$ times around the $3,4,\ldots,d$ directions.
Then a T-duality on the circles 1 and 2 will exchange 
the rank $N$ with the magnetic quantum number $M$. This is related to 
the N-duality transformation above (which exchanges the rank and 
electric quantum numbers) by S-duality. 

Concretely, in our case, we can consider the non-threshold electric
flux bound state of $N$ background D$d$-branes with a fundamental
NS string $\mbox{F1}_{d}$ wound $e_{d}$ times around the $d$'th circle. 
N-duality will map this to
\be
\begin{array}{lccc}
\mbox{Example I:} &N\;\mbox{D}d + e_{d}\; \mbox{F1}_{d}
&\stackrel{T^{d-1}}{\longrightarrow}& 
N\;\mbox{D1}_{d} + e_{d}\; \mbox{F1}_{d}\non
&&\stackrel{S_{IIB}}{\longrightarrow}& 
N\;\mbox{F1}_d + e_{d}\; \mbox{D1}_{d}\non
&&\stackrel{T^{d-1}}{\longrightarrow}& 
N\;\mbox{F1}_d + e_{d}\; \mbox{D}d\;\;,
\end{array}
\ee
and we thus see explicitly that it exchanges the quantum numbers 
$(e_{d},N)$, mapping the corresponding BPS states to each other.
This is reminiscent of the Nahm-duality transformation for
instantons on $T^{4}$ which provides an isomorphism between 
the rank $N$ instanton number $k$ and rank $k$ instanton number
$N$ instanton moduli spaces \cite{Nahm} (and precisely this
Nahm duality arises in the consideration of D$(p+4)$-D$p$ brane systems
\cite{douglasmoore,toronsetc}). 

Of course, changing $N$ also amounts to changing the longitudinal
momentum sector of M(atrix) theory described by its DLCQ. Thus,
N-duality actually relates BPS states corresponding to different
gauge groups and, as suggested in \cite{susskind}, the theories for
different values of $N$ should perhaps be combined into some larger
theory. As mentioned in the introduction, N-duality would then be a
signature of the Lorentz invariance of this theory.
Alternatively, we note that at least in the large $N$ limit,
many $U(N')$ BPS configurations for $N'<N$ are realized in the $U(N)$
theory itself (via reducible configurations), and this may be
relevant for the issue of Lorentz invariance in the original
BFSS \cite{bfss} matrix model. 

In fact, for $d=3$ one even
has a stronger statement. Namely, it is known \cite{Verlinde}
that the effective gauge group for a BPS configuration is $U(N')$ where
$N'\leq N$ is a U-duality invariant function of $N$ and the electric and
magnetic quantum numbers. For instance a magnetic flux configuration with
$N$ D$d$-branes and and an $M$-wrapped D$(d-2)$-brane (we will consider
the behaviour of this state under N-duality below), has effectively
a $U(N')$-theory where $N'=\mbox{gcd}(N,M)$ is the greatest common divisor
of $N$ and $M$. This can be seen by T-dualizing this to a D1-D1 string
system which effectively represents a single D1 string wrapping $N'$
times around a particular one-cycle of a two-torus. 

One can also see explicitly that, via the twisted (toron)
boundary conditions, the corresponding BPS state breaks the gauge symmetry 
down to  $U(N')$. It is in this sense that both the original $U(N)$
configuration
with quantum numbers $(N,M)$ and the new $U(M)$ configuration with 
$(N_{new}=M,M_{new}=N)$ are realized as BPS states in the same,
$U(N')$, gauge theory. If this property persists in some form in
higher dimensions, with all the relevant 
(five-brane, D-brane, \ldots) quantum
numbers included, N-duality could become a true symmetry of the BPS
spectrum of `SYM' theory, at least for large $N$.

\subsection{Problems with $N=0$}

In the above example, after the N-duality we again end up with a
configuration containing a top-dimensional brane (and thus permitting
a SYM-like interpretation in the standard sense) essentially because, 
on the M-theory side one is simply exchanging KK momentum quantum
numbers in the $d$'th and 11'th direction. However, in general this
need of course not be the case, as exemplified by the N-dual of the
magnetic flux state with quantum numbers $(N,m_{12})$ which, according
to  (\ref{ynq}) is mapped to a BPS state with zero D$d$-brane number,
magnetic quantum number $m_{12}$ and electric quantum number $e_{d}=N$.
Concretely, one has
\be
\begin{array}{lccc}
\mbox{Example II:}& N\;\mbox{D}d + m_{12}\; \mbox{D}(d-2)_{d}
&\stackrel{T^{d-1}}{\longrightarrow}& 
N\;\mbox{D1}_{d} + m_{12}\; \mbox{D3}_{d}\non
&&\stackrel{S_{IIB}}{\longrightarrow}& 
N\;\mbox{F1}_d + m_{12}\; \mbox{D3}_{d}\non
&&\stackrel{T^{d-1}}{\longrightarrow}& 
N\;\mbox{F1}_d + m_{12}\; \mbox{D}(d-2)_{d}\;\;,
\end{array}
\ee

Let us stress here that from the present
(passive) point of view (permutations
of quantum numbers) one is taking the configurations in the first line,
mapping them to those in the last line, but reinterpreting them (via
an analytic continuation in the parameters $(\gst,\ls,s_{i})$ defining
the \IIt\ theory) as configurations in the original \IIt\ theory 
It is this analytic continuation that is responsible for the
fact that U-duality relates BPS states with different energies. 

It is also responsible for the fact that, viewed this way, the action
of the full U-duality group apparently does not commute with taking
the matrix theory limit. In fact, in Example I one is genuinely
mapping a $U(N)$ flux state to a $U(e_{d})$ flux state, both of which
have well-defined finite matrix theory limits (albeit with respect to
different backgrounds) and both can be realized as $U(N')$ configurations
where $N'=\mbox{gcd}(N,e_{d})$. In Example II, on the other hand, one
finds that the well-defined magnetic flux bound state is mapped to a
non-threshold bound state with energy
\be
E=\sqrt{\left(\frac{Ns_{d}}{\ls^{2}}\right)^{2}
+\left(\frac{m_{12}V_{s}}{g^{2}\ls^{2}s_{i}s_{j}}\right)^{2}}\;\;,
\ee
whose mass diverges in the \IIt\ matrix theory limit $\ls\ra 0$
without there being the possibility to subtract a background
contribution while retaining a finite result.

The origin of this singular behaviour can be understood in a variety
of ways:
\begin{enumerate}
\item In the SYM matrix theory picture
this behaviour is clearly a consequence of the fact that the N-dual
configuration contains no top-dimensional wrapped D$d$-branes which
could serve as a background configuration for $(d+1)$-dimensional
SYM theory. 
In fact from the SYM$_{d+1}$ point of view such configurations,
like the above
F1-D$(d-2)$ system, are quite singular, describing 
distributional gauge field configurations where the gauge field 
is concentrated on a lower-dimensional cycle. In this sense it 
is not too surprising to find that its energy diverges in the `SYM'
matrix theory limit. 
\item From the IIA string theory point of view, the new
$E_{d+1}/E_{d}$ symmetries are associated with T-dualities
involving the light-cone direction. In \cite{cds} these 
have been shown to be quite singular in the absence of antisymmetric
tensor background fields (see also \cite{moore} for an extensive
discussion of time-like T-dualities).
\item Finally, from the DLCQ point of view
this may also simply be a dual manifestation
of the problems with zero light-cone momentum states in the DLCQ
(see \cite{mblfq} or \cite{shjp} for a recent discussion 
within the present context).
\end{enumerate}

We have not been able to resolve these problems. At first one may
have thought that this behaviour is an indication of the fact that
the U-duality group of a light-like compactification is smaller
than (or a contraction of) $E_{d+1}$, but the results of 
\cite{hulljulia} quoted in \cite{ch} suggest that this is not the case.
  
It has already been observed in the past that occasionally string
duality requires replacing the traditional gauge theory objects (vector
bundles) by something more general. In fact, in \cite{jhgm} it was
shown that compatibility of the analysis of 4-2-0 D-brane systems on
$T^{4}$ or a K3 with the predictions of string duality can be achieved
if one considers moduli spaces of simple coherent semistable (Chan-Paton)
sheaves rather than moduli spaces of vector bundles. In this setting,
brane configurations with and without top-dimensional branes, 4-branes
in this case, can be treated on an equal footing. In particular,
2-0 D-brane systems on a four-torus correspond to sheaves with
Mukai vector $(0,ch_{1},ch_{2})$ and these lead to well-defined
compact and smooth moduli spaces \cite{mukai}.

While this provides a satisfactory setting for discussing
D-brane dualities in $d=4$, it does not generalize immediately
in any obvious way to $d\neq 4$ and some new ideas appear to be required.
Very recently, the interesting proposal has been put forward by
Connes, Douglas and Schwarz \cite{cds} and Douglas and Hull \cite{mdch}
that the matrix theory limit of toroidal compactifications with
non-trivial background fields along the null direction is described
by SYM theory on a {\em non-commmutative} torus. It has been argued
in \cite{cds} that this deformation of SYM theory is necessary if one
wants to exhibit the full U-duality group expected from M-theory
in the matrix theory. 

As here we are mainly dealing with the Weyl subgroup of the U-duality
group, non-trivial background fields are not an issue (one can consistently
work with a rectangular torus and zero three-form field). Nevertheless,
the BPS mass formulae of \cite{cds} suggest that within the non-commutative
geometry setting also states with $N=0$ have well-defined finite energies
for a suitable class of modules. This might provide further evidence
in favour of the suggestion of \cite{cds,mdch} that non-commutative
geometry is a better arena for matrix theory than SYM on a commutative 
torus. It also raises the question 
as to whether there is some relationship, in $d=4$,
between non-commutative SYM theory and SYM theory for sheaves. 

\subsection{N-Duality, (1+1)-Dimensional Backgrounds and Matrix
Strings}

We have seen in the previous section that part of the U-duality
group is obscured in the matrix theory limit, when one
considers the passive action of the U-duality group on the BPS
states of a given string theory. On the other hand, as the group
$E_{d+1}$ is a manifestation of the general covariance of M-theory
compactified on a torus $T^{d+1}$, one expects a realization of the
full U-duality group to play a role in establishing the Lorentz
invariance of (some future reincarnation of) matrix theory. With this
in mind, in this section we shall focus on another realization of the
U-duality group, namely in its active sense.  In this interpretation, 
one is explicitly relating states in different string 
theories (e.g.\ T-duality changes the radii and exchanges 
type IIA with type IIB) which have the same energies. 

In the context of the first example above, for example, this means 
that one is now not dealing with a $U(e_{d})$ gauge theory in which 
the $e_{d}$ D$d$-branes are to be treated as background. Rather, after 
N-duality it is the image of the D$d$-brane, i.e.\ the F1 string, 
that is to be treated as the background field that becomes infinitely 
massive in the matrix theory limit. Indeed, the D$d$-brane mass in the
\IIt\ theory can be written as
\be
\frac{V_{s}}{g^{2}\ls^{4}}= \frac{s_{d}}{\ls^{2}\gamma_{N}}
\ee
where $\gamma_{N}$ was defined in (\ref{gammas}). Comparing with the
transformation rules (\ref{yn}) one sees that this is precisely the
mass of a fundamental string wrapped around the $d$-direction in the 
N-dual theory. U-duality in the active sense preserves masses and 
thus the F1 string is the background configuration. This (1+1)-dimensional
picture is guaranteed by the fact that 
N-duality not only always maps the 
top-dimensional D-brane to a wound F1 string (around the $d$-direction), 
but that also the other N-dual circles shrink to zero size in the matrix 
theory limit,
\be
s_{i} \ra \frac{\ls^{2}g^{2}s_{d}}{V_{s}}s_{i} 
\stackrel{\ls\ra 0}{\longrightarrow} 0
\;\;.
\ee
This means that there is always an effectively (1+1)-dimensional 
description of any bound state with a D$d$-brane, obtained by
Kaluza-Klein reduction. For example in the case of the
D$d$-D$(d-2)$-system, with the $(d-2)$-brane wrapping also the $d$-direction,
the D$(d-2)$-brane magnetic flux is represented
by a scalar field configuration on the world-sheet of the F1 string,
namely 
\be
\int \tr F_{ij} \ra \oint \tr [X_{i},X_{j}]\;\;,
\ee
where $X_{i}$ are the scalar fields corresponding to the components 
of the gauge field transverse to the string world-sheet. In fact, 
more generally, N-duality reproduces precisely the matrix string
flux-brane dictionary of \cite{dvv} in which the D$d$-brane number
of the \IIt\ theory (the D0-brane number of the IIA theory) is
represented by an electric flux on the string world-sheet and other
configurations correspondingly identified.

\subsection{An M2-brane Picture of Matrix Theory for $d$ Even}

Let us take a closer look at the matrix string picture we obtain
by acting with N-duality on the \IIt\ theory. The N-dual theory
is characterized by the string length $\lsh$ and the
string coupling constant $\gsh$. From (\ref{yn},\ref{yngs}) 
we have
\bea
\lsh^{2} &=& \frac{g^{2}\ls^{4}s_{d}}{\vs}\non
\gsh   &=& \frac{g^{d-3}}{\ls^{2}\vs^{(d-5)/2}}\;\;.
\eea
Thus $\lsh\ra 0$ as it should in the matrix theory limit. 
The string coupling constant $\gsh$, on the other hand, 
diverges and thus, as it stands, this matrix string picture
is perhaps not the most useful way of describing the situation.

At this point a peculiar distinction between $d$ even and $d$ odd 
arises.\footnote{This was noticed in a discussion with Tom Banks.} 
Namely, for $d$ odd, when one is in a type IIB string theory,
one can always perform an S-duality to arrive at a picture in which
both the new string length and the string coupling constant tend
to zero (as $\ls^{2}$) in the matrix theory limit. For $d=1$ this
reproduces exactly the standard (1+1)-dimensional SYM matrix 
theory on the D-string world-sheet. This is true rather trivially
as in $d=1$ N-duality is the same thing as the $S_{IIB}$-duality.
For $d=3$, on the other hand, this should
provide a well-defined D-string description of SYM theory in 
$(3+1)$-dimensions and it may be interesting to pursue this.

When $d$ is even, this procedure is not available. However, as one is
dealing with strongly coupled strings one is motivated to lift this
to an $\widehat{M}$-theory characterized by an $\rllh$ and a new
Planck length $\lph$. This indeed turns out to be a promising thing
to do as the length of the new eleventh dimension
is
\be
\rllh = \gsh\lsh = \frac{g^{d-2}s_{d}^{(d-4)/2}}{\vs^{(d-4)/2}}\;\;,
\ee
and is therefore constant in the matrix theory limit. Thus for
$d$ even the objects that appeared to be strings 
are actually M-theory membranes.
This picture is potentially useful due to the fact that the
new Planck length, 
\be
\lph^{3}=\lsh^{3}\gsh = \frac{g^{d}s_{d}^{(d-2)/2}\ls^{4}}{V_{s}^{(d-2)/2}}
\;\;,
\ee
goes to zero when $\ls\ra 0$, implying that the  
dynamics on the world-volume of this $\widehat{M}$-theory two-brane
decouples from the bulk dynamics in this limit. 

If this description is to be trusted, it should at the very least
reproduce the known matrix theory for $d=2$, the SYM$_{2+1}$-theory
on the world-volume of the D2-branes, and this is indeed the case.
To see this we note that for $d=2$ $\rllh=s_{1}$, and the new
radius of the 1-direction is $\widehat{s_{1}}\sim s_{1}\ls^{2}$. 
Thus the $\widehat{M}2$-brane becomes a D2-brane 
in the new IIA string theory associated with shrinking the 1-direction
of the $\widehat{M}$-theory, and it is wrapped around the (constant)
2- and 11-directions.

Something more interesting appears to happen for $d=4$. In that case,
as we recalled in section 2.3, the matrix theory is actually the
world-volume theory on the M5-brane of the $\widetilde{M}$-theory
associated to the \IIt\ string theory \cite{rozali,seiberg45}. As the
M5-brane is the electro-magentic dual of the M2-brane, we would
like to suggest that the above description is precisely a 
dual description of the known matrix theory for $d=4$. In favour
of this interpretation, which certainly needs to be substantiated,
we note that the parameters of the $\widehat{M}$- and 
$\widetilde{M}$-theories are related by
\bea
\rllh &=& \lsh\gsh = g^{2} = \gst\ls =\widetilde{R_{11}}\non
\lph^{3} &=& \widetilde{\ell_{P}}^{6}\frac{s_{d}}{V_{s}}\;\;.
\eea
This is precisely as required by M-theory 2-brane/5-brane duality
which does not change $R_{11}$ but exchanges the M2- and M5-brane
tensions $\ell_{P}^{3}$ and $\ell_{P}^{6}$ respectively.

Naively at least, this membrane picture appears to be valid
also for $d=6$, thus prompting the suggestion
that this may provide a way of understanding the elusive
matrix theory for $d=6$.

Notice that in the (2,1) string construction of M-theory \cite{km}, 
depending upon
the choice of null gauging, one finds either the world-volume
theory of a D-string or of a D2-brane as the target space of the
(2,1) string worldsheet. 
These are also the only two possibilities that arise in the
constructions described above. In \cite{martinec} evidence was 
presented that the (2,1) string theory is related to 
the maximally compactified matrix theory. Our observation that in all 
the constructions of lower-dimensional world-volume theories
obtained by the action of N-duality on a BPS configuration 
of the \IIt\ theory one only obtains either string or membrane
pictures, just as in the (2,1) string constructions, provides 
further evidence in support of this conjecture. 

\subsection{The Invariant $I$ and Black-Hole Entropy}

Let us briefly come back to the U-duality invariant $I_{d}$
(\ref{minv},\ref{yinv}) 
which was important for the algebraic analysis of section 3.
For the present purposes we will find it convenient to
rewrite this in the compact form
\be 
I_{d}^{-1} = \frac{V_{s}}{\gst^{2}\ls^{8}} 
\label{yinv2}
\ee
valid for any $d$. Although perhaps not immediately obvious
at this point, this is precisely the invariant that plays 
a prominent role in the analysis of supergravity BPS states
and black hole entropies (see \cite{fm,khhms}). In fact, we
will now show that for $d=5,6$, corresponding to black holes
in five and four dimensions, $I$ coincides with the cubic
invariant of $E_{6}$ and the quartic invariant of $E_{7}$
respectively.

In the construction of Strominger and Vafa \cite{stromvaf}
five-dimensional black holes are labelled by their
D5-brane number $Q_{5}$, as well as by the D-string winding
number $Q_{1}$ and momentum $P$ in one (and the same)
direction. The product of the three dressed charges (fluxes, tensions)
is precisely
\be
\frac{Q_{5}V_{s}}{\gst\ls^{6}} . \frac{Q_{1}s_{6}}{\gst\ls^{2}} .
\frac{P}{s_{6}} = (Q_{5}Q_{1}P)I_{5}^{-1}\;\;.
\ee
Alternatively the black holes may be described in the S-dual 
basis of NS5-brane number, string winding and momentum, and of course
one finds
\be
\frac{Q_{5}V_{s}}{\gst^{2}\ls^{6}} . \frac{Q_{1}s_{6}}{\ls^{2}} .
\frac{P}{s_{6}} = (Q_{5}Q_{1}P)I_{5}^{-1}\;\;,
\ee
as guaranteed by the U-duality $E_{6}$-invariance of $I_{5}$.

Four-dimensional black holes, corresponding here
to $d=6$, are labelled by four parameters \cite{stromal}
and their entropy
is given in terms of a quartic invariant of $E_{7}$. $I_{6}^{-1}$
has mass-dimension 2, so we expect this quartic invariant to be
related to $I_{6}^{-2}$. This is indeed the case. For example,
a BPS black hole configuration can be labelled by four D3-brane
winding numbers corresponding to D3-branes  wrapping the (123),
(345), (561), (246) directions. Computing the product of
their tensions (as above we could include the corresponding
quantum numbers), one finds
\be
\frac{s_{1}s_{2}s_{3}}{\gst\ls^{4}} .
\frac{s_{3}s_{4}s_{5}}{\gst\ls^{4}} .
\frac{s_{5}s_{6}s_{1}}{\gst\ls^{4}} .
\frac{s_{2}s_{4}s_{6}}{\gst\ls^{4}} 
= \left(\frac{V_{s}}{\gst^{2}\ls^{8}}\right)^{2} = I_{6}^{-2}
\ee
Alternatively, a four-dimensional black hole can be described
by the D6-brane number and three mutually orthogonal D2-brane
winding numbers, leading to the expected result
\be
\frac{V_{s}}{\gst\ls^{7}} . \frac{V_{s}}{(\gst\ls^{3})^{3}} = I_{6}^{-2}
\;\;.
\ee
This agreement between the U-invariant (\ref{yinv2}) and the 
invariants appearing in the discussion of black hole entropies
is, of course, virtually guaranteed by the paucity of $E_{d+1}$
invariants. But from the traditional point of view the cubic
invariant of $E_{6}$ and the quartic invariant of $E_{7}$ appear
to be very different objects. (\ref{yinv2}), on the other hand,
provides a general and unified expression for the U-duality 
$E_{d+1}$-invariant for any $d$.

\subsection{A Look at Some States in $d=6$ and $d=7$}

We have already mentioned in section 3.4 that in $d\leq 5$ only one new
state, the background $\N$, is needed to unify the momentum and flux
$E_{d}$-multiplets into one $E_{d+1}$-multiplet (in the
sense of one-particle states). In $d\geq 6$ however, more new states
necessarily appear and these should somehow be indicative of the
new physics that appears in the matrix theory description of
lower-dimensional string compactifications.

For example, for $d=6$, instead of 
the $27+27=54$ flux and momentum states in the $\rb{27}\oplus\r{27}$
of $E_{6}$, we find $56$ states in the Weyl group orbit of the
$\r{56}$ of $E_{7}$. The new state arises by considering the 
$SL(7,\ZZ)$ 7-plet corresponding to the Taub-NUT TN6-brane on the
M/IIA-theory side, with mass
\be
M=\frac{VR_{a}}{\ell_{P}^{9}}\;\;,\;\;\;\;\;\;a=1,\ldots,6,11\;\;.
\ee
Here and below $V$ denotes the $(d+1)$-dimensional volume $V=V_{R}R_{11}$.
Choosing $R_{a}$ to be one of the spatial circles,say $R_{6}$,
one obtains the TN5-brane, i.e.\ the KK monopole, with $R_{6}$ the NUT
direction. This turns into the NS5-brane (transverse to $R_{6}$) in the 
\IIt\ theory, with mass
\be
M=\frac{VR_{6}}{\ell_{P}^{9}}=\frac{V_{s}}{s_{6}\gst^{2}\ls^{6}}\;\;.
\label{d6ns5}
\ee
For $R_{a}=R_{11}$, on the other hand, one obtains
the D6-brane, thus a D0-brane in the \IIt\ theory, with mass
\be
M=\frac{VR_{11}}{\ell_{P}^{9}}=\frac{1}{\gst\ls}\;\;.
\ee
Clearly these two states are related by N-duality, using $\a_{d=6}$, as can
also be checked directly by acting with $T_{12345}S_{IIB}T_{12345}$. 

The D0-brane does not appear in the SYM flux and
momentum multiplets as it cannot form a BPS bound-state with the
background D6-branes defining the low-energy SYM theory. In fact, we
have to remember that N-duality also acts non-trivially on the background,
so not all single-particle states in the orbit of the U-duality group need
necessarily be able to form (or appear as) BPS bound states with the
background D$d$-branes. 

However, just using S-dualities (and permutations),
one can e.g.\ map the threshold bound state formed by the background $N$
D6-branes 
and wrapped D2-branes (instantons) to a threshold bound state consisting of an
NS5-brane wrapped around the background D6-branes. This state is then
of course well-defined in the matrix theory limit (note that (\ref{d6ns5})
is constant in that limit as $g^{2}=\gst\ls^{3}$ in $d=6$), and so is
therefore its N-dual. As N-duality maps the D6 branes to wrapped NS strings,
this N-dual bound state is thus a BPS threshold bound state of D0-branes
with $N$ background NS strings wrapped around the 6-direction, once
again providing a (1+1)- or (2+1)-dimensional picture of this configuration. 

Let  us now consider $d=7$. Instead of 
the $56+126=182$ flux and momentum states in the $\r{56}\oplus\r{133}$
of $E_{7}$ we find $240$ states in the Weyl group orbit of the
$\r{248}$ of $E_{8}$. Their M-theory masses and those of their 
\IIt\ counterparts, together
with their $SL(8,\ZZ)$ degeneracy, are displayed in the following table.
There $R_{a}, a =1,\ldots,7,11$ denotes one of the $(d+1)=8$ radii of
M-theory, and we have
divided the \IIt\ states into longitudinal (L) and transverse (T)
states according to whether one of the $R_{a}$ is $R_{11}$ or not
(this is for book-keeping purposes only - it does not 
mean that these states are necessarily longitudinal or transverse on the 
M-theory side).
\be
\begin{array}{|c|c|c|c|}\hline
SL(8,\ZZ)& \mbox{M-Theory} & \mbox{\IIt\ (T)}&\mbox{\IIt\ (L)}\\ \hline
8 & \frac{1}{R_{a}} & \frac{s_{i}}{\ls^{2}}& \frac{V_{s}}{g^{2}\ls^{4}} 
\\ \hline
28 & \frac{R_{a}R_{b}}{\ell_{P}^{3}} & \frac{V_{s}}{g^{2}\ls^{2}s_{i}s_{j}}
& \frac{1}{s_{i}}\\ \hline
56 & \frac{V}{\ell_{P}^{6} R_{a}R_bR_c} & \frac{s_{i}s_{j}s_{k}}{g^{2}}
& \frac{V_{s}s_{i}s_{j}}{g^{4}\ls^{2}}\\ \hline 
56 & \frac{V R_a}{\ell_{P}^{9} R_{b}} & \frac{V_{s} s_i}{g^{4}s_j} &
\frac{V_{s}^{2}}{g^{6}\ls^{2}s_{i}}\& \frac{s_{i}\ls^{2}}{g^{2}} \\ \hline
56 & \frac{VR_aR_bR_c}{\ell_{P}^{12}} & \frac{V_{s}^{2}}{g^{6}s_{i}s_j s_k} &
\frac{V_{s}\ls^{2}}{g^{4}s_{i}s_{j}} \\ \hline
28 & \frac{V^{2}}{R_{a}R_{b}\ell_{P}^{15}} &
\frac{V_{s}\ls^{2}s_{i}s_{j}}{g^{6}} & \frac{V_{s}^{2}s_{i}}{g^{8}}
\\ \hline
8 & \frac{V^{2}R_{a}}{\ell_{P}^{18}} & \frac{V_{s}^{2}\ls^{2}}{g^{8}s_{i}}
& \frac{V_{s}\ls^{4}}{g^{6}} \\ \hline
\end{array}
\label{tabled7}
\ee
As little (or practically nothing) appears to be known about matrix
theory on $T^{7}$ (but see \cite{ch}), 
a concrete (or even tentative) identification of some
of the states on the `SYM' side will be difficult to come by. However,
let us make a few comments on this U-duality spectrum:

\begin{enumerate}
\item
To understand at least in part, some of the new states appearing
in this table, recall that carrying out a T-duality on a circle
transverse to the world volume of the NS5-brane produces 
a TN5-brane and vice versa. Once we compactify on a 7-torus 
the NS5-brane has two transverse directions and 
we can apply two independent transverse T-dualities. 
{}From the NS5-brane of the \IIt\ theory, two T-dualities
produce the configuration with mass 
$\frac{V_{s}s_{i}s_{j}}{g^{4}\ls^{2}}$. 

\item The states in this 
table that really require some interpretation however are
not these, but rather those with a mass behaving as inverse
powers of $g^2$ or $\gst$ greater than $2$. 
For instance applying N-duality to the non-NUT transverse direction of the
TN5-brane, one obtains a `rolled-up' version of the TN6-brane, 
corresponding
to the first set of longitudinal states in the fourth row. Such a
configuration has been considered
(in a very different context) in \cite{ovnsss} and will be discussed
in more detail in the next section. Note that this latter set of states
displays the unusual $1/g_{s}^{3}$-dependence on the IIA/M-theory side
noted in \cite{egkr}. 
The \IIt-counterpart of this 
rolled-up TN6-brane, which displays a $1/\gst^{3}$-behaviour,
can form a half-BPS bound state with the background D7-branes with energy
$E_{SYM}=V_{s}^{3}/Ng^{10}s_{i}^{2}$. Note that this is just one
of many states in the table that have this peculiar 
dependence on the coupling constants. 

\item In total the new states we find are the 
longitudinal state in the first row,
seven longitudinal states in the fourth row (last entry), twenty-one
longitudinal states  in the fifth row, twenty-one transverse states in
the sixth row, and finally all of the eight states in the last row.

\item  
The new states in the fourth row can be concretely identified as wound
D1-strings since $s_{i}\ls^{2}/g^{2} = s_{i}/\gst\ls^{2}$. Although
of course wound D1-strings can appear in every odd dimension, these
play a special role here for the same reason that D0-branes are
special in $d=6$: they cannot form BPS bound states with the background
D$d$-branes and thus they can only appear after a transformation like
N-duality that acts non-trivially on the background. 
\end{enumerate}

\subsection{A Rolled-Up Taub-NUT Soliton and $1/g_{s}^{3}$-States}

For compactifications with $d\geq 7$, it was observed in \cite{egkr}
that in the U-duality multiplet one finds states which in the 
IIA string theory picture have masses that behave like
$1/g_s^3$ or higher inverse powers of $g_s$. These are clearly neither 
D-brane nor solitonic p-brane objects. For objects of this type 
it was argued in \cite{egkr} that the gravitational field 
will be very large - in particular they will not 
correspond to space-times that are asymptotically flat. One can 
actually construct an explicit example of such a metric by
considering for $d=7$ the states with mass
\be
M = {R_a^2 R_{b_1}\ldots R_{b_6}\over l_P^9},
\label{TN6}
\ee
This is the Taub-NUT 6-brane of M-theory. Reaching the 
IIA string by compactifying on the $R_a$ circle leads
to the D6-brane with mass proportional to the 
product of the 6 radii divided by $g_sl_s^7$; compactifying 
on one of the $R_{b_i}$ we find the TN5-brane of IIA string theory,
with mass proportional to a product of radii (with one squared 
as for the TN6-brane) divided by $g_s^2l_s^8$. However, if we compactify
on the remaining one of the $d+1=8$ circles of M-theory we obtain an 
object with mass
\be
M = {R_a^2 R_{b_1}\ldots R_{b_6}\over g_s^3l_s^9}.
\label{newb}
\ee

Given that we started with a TN6-brane, this compactification
clearly corresponds to taking an infinite array 
of TN6-branes and compactifying along this periodic direction.

The multi-TN metric has the form
\bea
ds^2 &=& V^{-1} (d\psi + \vec{A}.d\vec{x})^2 + V d\vec{x}^2\non
V &=& {1\over L^2} + {1\over 2}\sum_{i=1}^k {\ell_P\over |\vec{x} - 
\vec{x}_i|}\;\;,
\eea
$\psi \in[0, 2\pi \ell_P], \vec{x} = (x,y,z)$. The length of the circle at
$\infty$ is $2\pi \ell_PL$. 

Now consider a configuration with $\vec{x}_i = (ia,0,0)$ for $i\in \ZZ$. 
Note that the 
sum in $V$ now diverges, but subtracting an infinite constant
we can perform the sum using Poisson resummation. The result is 
a new metric with $V$ given by,
\be
V(x) = {1\over 2\pi} \log ({\mu a\over \rho}) + 
\sum_{m\neq0} \ex{2\pi im{x\over a}} K_0({2\pi|m|\rho\over \lambda a}),
\ee
where  $\mu$ is some constant and $\rho = \sqrt{y^2 + z^2}$. The 
new metric is periodic in $x$ and $\psi$ and choosing the $R_{11}$ direction 
to correspond to $x$ we find the metric corresponding to the 
state with mass proportional to $1/g_s^3$. It is amusing to notice that 
precisely this metric arises in the analysis 
of the hypermutiplet moduli space of the IIA string theory in 
the vicinity of the conifold  singularity \cite{ovnsss}. Furthermore, due
to the explicit appearance of the logarithmic term in V, the 
metric is not asymptotically flat but $\log$-divergent, as predicted by 
the  general arguments of \cite{bs}. 

In $d=7$ as we saw in the previous subsection, 
there are many states of this type with mass proportional 
to higher inverse powers of $g$, E.g.\ the M-theory mass of the 
states in the fifth row is, 
\be
M=\frac{R_{a}^{2}R_{b}^{2}R_{c}^{2}R_{d_1}R_{d_2}R_{d_3}R_{d_4}R_{d_5}}%
{\ell_{P}^{12}}\;\;.
\ee
This seems to require some $5d$ Euclidean configuration 
(gravitational instanton) with three NUT directions,
generalizing the Euclidean Taub-NUT solution in $d=4$.
Such an object appears not to be known. U-duality, however, would
predict that it can form a threshold bound state with D7-branes.

\subsubsection*{Acknowledgements}

We are grateful to Tom Banks, Fawad Hassan, K.S.\ Narain and George
Thompson for useful discussions at various stages of this work. This 
work was supported in part by the EC under 
the TMR contract ERBFMRX-CT96-0090. 

\rnc{\Large}{\normalsize}

\end{document}